\documentclass[twocolumn,preprintnumbers,amsmath,superscriptaddress,amssymb,aps] {revtex4}
\usepackage{graphicx}
\usepackage{dcolumn}
\usepackage{bm}
\def\degree{${}^{\circ}$}

\begin{document}
%\title{Magnetism and Magneto-optical Kerr effect in double perovskite Ba$_2$NiOsO$_6$ and
%its (111) (Ba$_{2}$NiOsO$_{6}$)/(BaTiO$_{3}$)$_{10}$ superlattice from first principles calculations}
%\title{Large magneto-optical effects in ferromagnetic topological semiconductors Ba$_2$NiOsO$_6$ and
%(Ba$_{2}$NiOsO$_{6}$)/(BaTiO$_{3}$)$_{10}$ (111) bilayer: First principles calculations}
%\title{First-principles study of magneto-optical effects in ferromagnetic topological semiconductors
%Ba$_2$NiOsO$_6$ and (Ba$_{2}$NiOsO$_{6}$)/(BaTiO$_{3}$)$_{10}$ (111) bilayer}
\title{Anomalous ferromagnetism and magneto-optical Kerr effect in semiconducting double perovskite Ba$_2$NiOsO$_6$ and
its (111) (Ba$_{2}$NiOsO$_{6}$)/(BaTiO$_{3}$)$_{10}$ superlattice}
\par
\author{Hai-Shuang Lu}
\address{Department of Physics and Center for Theoretical Physics, National Taiwan University, Taipei 10617, Taiwan}
\address{College of Physics and Electronic Engineering, Changshu Institute of Technology, Changshu 215500, People's Republic of China}
\address{Physics Division, National Center for Theoretical Sciences, Hsinchu 30013, Taiwan}
\author{Guang-Yu Guo}
\email{gyguo@phys.ntu.edu.tw}
\address{Department of Physics and Center for Theoretical Physics, National Taiwan University, Taipei 10617, Taiwan}
\address{Physics Division, National Center for Theoretical Sciences, Hsinchu 30013, Taiwan}
\date{\today}

\begin{abstract}
We carry out a first-principles investigation on magnetism, electronic structure, magneto-optical effects
and topological property of newly grown cubic double perovskite Ba$_{2}$NiOsO$_{6}$ and its (111)
(Ba$_{2}$NiOsO$_{6}$)/(BaTiO$_{3}$)$_{10}$ superlattice, based on the density functional theory
with the generalized gradient approximation (GGA) plus onsite Couloumb interactions. % in the GGA + $U$ scheme.
Interestingly, we find that both structures are rare ferromagnetic (FM) semiconductors with estimated Curie temperatures
of $\sim$150 and 70 K, respectively.
The calculated near-neighbor exchange coupling parameters reveal that
the ferromagnetism is driven by exotic FM coupling between Ni and Os atoms, which is due to
the FM superexchange interaction caused by the abnormally strong
hybridization between the half-filled Ni $e_{g}$ and unoccupied Os $e_{g}$
orbitals. The strong spin-orbit coupling (SOC) on the Os atom
is found to not only open the semiconducting gap but also produce a large antiparallel orbital
magnetic moment on the Os atom, thus reducing the total magnetization
from 4.0 $\mu_B$/f.u., expected from the Ni$^{2+}$ 3$d^8$
($\emph{t}_{2g}^{6}$ $\emph{e}_{g}^{2}$; $S=1$) and Os$^{6+}$
5$d^{2}$($\emph{t}_{2g}^{2}$ $\emph{e}_{g}^{0}$; $S=1$) ions.
%to the measured one of 3.37 $\mu_B$/f.u.
Remarkably, we also find that because of the enhanced exchange interaction on
the Os atoms caused by the Ni 3$d$ - Os 5$d$ hybridization and the strong
SOC of the Os atoms, the magneto-optical (MO) effects are large in these two structures.
For example, the Kerr and Faraday rotations in bulk Ba$_{2}$NiOsO$_{6}$ can be
reach 6$^{\circ}$ and 250 deg/$\mu$m, respectively, which are larger than
that of best-known MO materials. These interesting findings thus suggest that
because of their FM semiconductivity and excellent MO properties,
both structures would be promising materials for not only semiconductor-based spintronics
but also magneto-optical devices.
Finally, our calculated anomalous Hall conductivity shows that the band gap
just below the Fermi level in the superlattice is topologically nontrivial with the gap
Chern number of 2. This indicates that the (111) Ba$_{2}$NiOsO$_{6}$ and related 5$d$ double-perovskite
monolayers may provide an interesting material platform
for exploring magnetic topological phases and phase transitions.

\end{abstract}

\maketitle

\section{Introduction}
Double perovskite oxides A$_{2}$BB$'$O$_{6}$ (A is a
rare-earth/alkaline-earth cation; B and B$'$ are transition metal
cations), discovered in 1960s~\cite{Sleight}, have attracted
enormous attention in the past decades. They have been found to
show diverse properties, such as large room-temperature
magnetoresistance~\cite{Kobayashi98},
multiferroicity~\cite{Azuma05},
half-metallicity~\cite{Kobayashi98,Jeng03,Philipp003}, and
magneto-optic (MO) effects~\cite{vidya2004huge,Das08}, depending
on the compositions of the A, B, and B$'$ cations. More recently,
monolayers and multilayers of these double perovskites containing
heavy cations were predicted to host various topological
insulating phases such as quantum anomalous Hall
phase~\cite{Cook14,Cook16,Zhang14,Dong16}. Therefore, double
perovskite oxides offer ample possibilities for exploration of
spin-related physics and also for magnetic, magneto-electric and
MO device applications.

Recently, Feng {\it et al.}~\cite{Feng16} synthesized new double
perovskite Ba$_{2}$NiOsO$_{6}$ and found it to be a rare
ferromagnetic (FM) semiconductor with Curie temperature $T_{C}$ =
100 K~\cite{Feng16}. It crystallizes in a cubic $Fm\bar{3}m$
structure with lattice constant $a = 8.0428$ \AA, where the
Ni$^{2+}$ and Os$^{6+}$ ions are perfectly ordered on the B and
B$'$ sites, respectively~\cite{Feng16}. This is interesting
because FM semiconductors are rare and also useful for the
development of spintronic devices. Surprisingly, we note that Ni
and Os ions are ferromagnetically coupled, which is very rare
between the B and B$'$ cations in double perovskite
oxides\cite{Wang06,Wang09}. Furthermore, first-principles
electronic structure calculations~\cite{Feng16} showed that the
spin-orbit coupling (SOC) of Os$^{6+}$ plays a crucial role in
opening the semiconducting gap, and thus double perovskite
Ba$_{2}$NiOsO$_{6}$ is called a Dirac-Mott insulator. However, it
was inferred from the x-ray absorption spectra (XAS)~\cite{Feng16}
that the formal electronic configurations for Ni and Os in
Ba$_{2}$NiOsO$_{6}$ are Ni$^{2+}$ 3$d$$^{8}$($\emph{t}_{2g}^{6}$
$\emph{e}_{g}^{2}$; $S=1$; 2 $\mu$$_{B}$) and Os$^{6+}$
5$d$$^{2}$($\emph{t}_{2g}^{2}$ $\emph{e}_{g}^{0}$; $S=1$; 2
$\mu$$_{B}$), respectively. Consequently, the total moment should
be 4 $\mu$$_{B}$/f.u. for the FM ground state. However, the
measured saturation magnetization for Ba$_{2}$NiOsO$_{6}$ is
approximately 3.46 $\mu$$_{B}$/f.u. at 5 K and 50 kOe, much
smaller than the spin only magnetic moment estimated from the
simplified ionic model. Therefore, it would be interesting to
study the origin of the abnormal ferromagnetism as well as the
spin and orbital magnetic moments of Ba$_{2}$NiOsO$_{6}$.

When a linearly polarized light beam hits a magnetic material, the
polarization vector of the reflected and transmitted light beams
rotates. The former and latter are known as Kerr and Faraday
effects, respectively. Discovered in the 19th Century, they are
two well-known MO effects~\cite{Oppeneer2001,Antonov2004}.
Currently, MO Kerr effect (MOKE) is widely used as a powerful
probe of the electronic and magnetic properties of materials, such
as two-dimensional ferromagnetic
order~\cite{Gong2017,Huang2017,Fang2018}, spin Hall
effect~\cite{Stamm17}, skyrmion Hall effect\cite{Jiang17},
magnetic anisotropy~\cite{Lehnert10,HeLY15}, and topological
insulator~\cite{MacDonald10,Armitage12}. Furthermore, because of
its applications in modern high-density data-storage
technology~\cite{Mansuripur95}, an enormous amount of effort has
been devoted to search for materials with large MO signals.

Band exchange splitting caused by magnetization together with
relativistic SOC has been recognized as the origin of
MOKE~\cite{Oppeneer2001,Antonov2004,Argyres55,Erskine73a,Erskine73b}.
Localized 3$d$ orbitals tend to have large band exchange
splittings. However, their SOC is weak. However, 4$d$ or 5$d$
transition metal atoms have a strong SOC. Nonetheless, their
intra-atomic exchange interaction is rather small because of their
more extended $d$ orbitals which result in small band exchange
splittings. Therefore, an effective way to enhance the MOKE is to
make alloys or multilayers of 3$d$ transition metals with 4$d$ or
5$d$ transition metals\cite{Guo96}. Consequently, the
magneto-optical properties of various 3$d$ FM transition metal
alloys and multilayers with heavier 4$d$ or 5$d$ transition metal
atoms have been investigated extensively. For example,
PtMnSb~\cite{van Engen83} has been found to be an excellent MO
metal with a maximum Kerr rotation of 2.5$^{\circ}$. Double
perovskites, A$_{2}$BB$'$O$_{6}$ (B = 3$d$ and B$'$ = 4$d$ or 5$d$
transition metal elements), which can establish an unusual
renormalization of the intra-atomic exchange strength to enhance
the band exchange splitting at the 4$d$ or 5$d$ sites arising from
the so called hybridization driven
mechanism~\cite{Sarma000,Kanamori001}, is also expected to have
large MOKE. However, the B and B$'$ atoms in most of double
perovskite materials~\cite{Jeng03,Wang06,Wang09} prefer an
antiferromagnetic coupling. This may reduce the net magnetization
and thus results in a small MO effect~\cite{Das08}. As mentioned
above, simultaneous occurrence of ferromagnetism and
semiconducting gap in double perovskites is very rare. Combination
of the strong SOC of Os atoms and ferromagnetism thus would make
Ba$_{2}$NiOsO$_{6}$ an excellent semiconductor for not only
semiconductor-based spintronics but also magneto-optical devices.

The recent development in synthesizing artificial
atomic-scale transition metal oxide heterostructures provides
great tunability over fundamental physical parameters to realize
novel properties and functionalities~\cite{Mannhart010,Hwang12},
such as the conductive interface between two insulating oxides
~\cite{Ohtomo04,Lee2016,Lu15}. This also stimulates extensive
investigations on the topology of the electronic band structure of
transition metal oxide heterostructures~\cite{Xiao11,Cook14,Zhang14,Dong16,Chandra17,Hslu2018}.
Indeed, the quantum anomalous Hall phase was predicted to occur in (001)
double-perovskite La$_{2}$MnIrO$_{6}$ monolayer~\cite{Zhang14} and
also (111) double-perovskite La$_{2}$FeMoO$_{6}$ and
Ba$_{2}$FeReO$_{6}$ monolayers~\cite{Cook14,Baidya16}.
Therefore, it would also be interesting to study the topological properties of
Dirac-Mott semiconductor Ba$_{2}$NiOsO$_{6}$ and its heterostructures.

In this paper, we present a systematic first-principles study of
magnetism, electronic structure, magneto-optical effects and
topological property of cubic double perovskite
Ba$_{2}$NiOsO$_{6}$ and its (111)
(Ba$_{2}$NiOsO$_{6}$)$_{1}$/(BaTiO$_{3}$)$_{10}$ monolayer
superlattice. First, we find that both structures are narrow band
gap FM semiconductors. The ferromagnetism is driven by the rare
nearest-neighbor FM coupling between Ni and Os atoms, which is
attributed to the FM superexchange mechanism caused by the
abnormally strong hybridization between the half-filled Ni $e_{g}$
and unoccupied Os $e_{g}$ orbitals. Second, the strong SOC on the
Os atom is found to not only open the semiconducting gap but also
give rise to a large negative orbital magnetic moment on the Os
atom, thus resulting in a measured total magnetic moment of less
than 4 $\mu_B$/f.u.~\cite{Feng16}. Third, we also find that
because of the enhanced intra-atomic exchange interaction on the
Os atoms caused by the Ni 3$d$ - Os 5$d$ hybridization and the
strong SOC on the Os site, the MO effects are large in these two
structures. Our theoretical findings thus suggest that double
perovskite Ba$_{2}$NiOsO$_{6}$ and its (111) superlattice are
valuable ferromagnetic semiconductors for not only
semiconductor-based spintronics but also magneto-optical devices.
Finally, our calculated anomalous Hall conductivity reveals that
the band gap just below the Fermi level in the superlattice is
topologically nontrivial with the gap Chern number of 2. This
indicates that the (111) Ba$_{2}$NiOsO$_{6}$ monolayer
superlattice and related 5$d$ double-perovskite monolayers may
provide an interesting material platform for exploring magnetic
topological phases and phase transitions.

\section{THEORY AND COMPUTATIONAL DETAILS}

We consider cubic double perovskite Ba$_{2}$NiOsO$_{6}$ [Fig.
1(a)] and its (Ba$_{2}$NiOsO$_{6}$)$_{1}$/(BaTiO$_{3}$)$_{10}$
monolayer superlattice grown along the [111] direction [Figs. 1(b)
and 1(c)]. The Brillouin Zone (BZs) of bulk Ba$_2$NiOsO$_6$ and
the (111) (Ba$_2$NiOsO$_6$)$_{1}$/(Ba$_2$TiO$_3$)$_{10}$
superlattice are shown in Figs. 1(e) and 1(f), respectively.
Clearly, the latter is the folded BZ of the former along the
$\Gamma$ - L direction. In the present calculations of the
electronic structure and magneto-optical properties of bulk
Ba$_{2}$NiOsO$_{6}$, the experimental lattice constant of 8.0428
\AA $_{}$ is adopted. Since the BaTiO$_{3}$ slab in the
(Ba$_{2}$NiOsO$_{6}$)$_{1}$/(BaTiO$_{3}$)$_{10}$ superlattice is
much thicker than the Ba$_2$NiOsO$_6$ layer, the BaTiO$_{3}$ slab
could be regarded as the substrate. Therefore, the in-plane
lattice constant is fixed at ${\sqrt{2}}$${a_{0}}$ = 5.6962 \AA,
where ${a_{0}}$ is the theoretically determined lattice constant
of cubic perovskite BaTiO$_3$. In our structural optimization, the
in-plane hexagonal symmetry is fixed but lattice constant c and
internal coordinates of all the atoms in the superlattice are
optimized theoretically. The lattice parameters and atom positions
for bulk Ba$_{2}$NiOsO$_{6}$ and its (111) monolayer superlattice
are given, respectively, in Tables S1 and S2 in the Supplementary
Materials (SM)~\cite{See Supplemental}. The electronic structure
and magnetic structure are calculated based on the density
functional theory (DFT) with the generalized gradient
approximation (GGA)~\cite{Perdew96}. The accurate
projector-augmented wave (PAW) method~\cite{PEB}, as implemented
in the Vienna {\it ab initio} simulation package
(VASP)~\cite{Kresse93}, is used. The relativistic PAW potentials
are adopted to include the SOC. The valence configurations of Ba,
Ni, Os, Ti and O atoms adopted in the present calculations are
5\emph{s}$^{2}$5\emph{p}$^{6}$6\emph{s}$^{2}$,
3\emph{p}$^{6}$3\emph{d}$^{8}$4\emph{s}$^{2}$,
5\emph{p}$^{6}$5\emph{d}$^{6}$6\emph{s}$^{2}$,
3\emph{s}$^{2}$3\emph{p}$^{6}$3\emph{d}$^{2}$4\emph{s}$^{2}$ and
2\emph{s}$^{2}$2\emph{p}$^{4}$, respectively. To better account
for the on-site electron correlation on the \emph{d} shells of Os
and Ni atoms, the GGA + U method~\cite{dudarev98} is adopted with
the effective Coulomb repulsion energy $U_{Os}$ = 2.0 eV and
$U_{Ni}$ = 5.0 eV, respectively. The results of test calculations
using different $U$ values of 3 $\sim$ 5 eV for Ni and 1 $\sim$ 3
eV for Os, are given in Appendix A below. A large plane-wave
cutoff of 450 eV and the small total energy convergence criterion
of 10$^{-5}$ eV are used throughout. Fine Monkhorst-Pack
\emph{k}-meshes of 30$\times$30$\times$30 and
10$\times$10$\times$2 are used for the bulk and superlattice
calculations, respectively.

To find the ground state magnetic configuration and to understand
the magnetic interactions in both systems, we consider four
possible magnetic structures, as labeled FM, AFM, FI1, and FI2 in
Figs. 2 and 3. One can then evaluate the nearest-neighbor Ni-Os
(\emph{J$_{1}$}), Os-Os (\emph{J$_{2}$}) and Ni-Ni
(\emph{J$_{3}$}) magnetic coupling parameters by mapping the
calculated total energies of the FM, AFM, FI1, and FI2 magnetic
configurations to the classical Heisenberg model \emph{H} =
\emph{E$_{0}$} -
$\sum$$_{i>j}$\emph{J$_{ij}$}($\hat{e}_{i}\cdot\hat{e}_{j}$),
where \emph{J$_{ij}$} is the exchange coupling parameter between
sites \emph{i} and \emph{j}, and $\hat{e}_{i}$ denotes the
direction of spin on site \emph{i}.

\begin{figure}
\includegraphics[width=8cm]{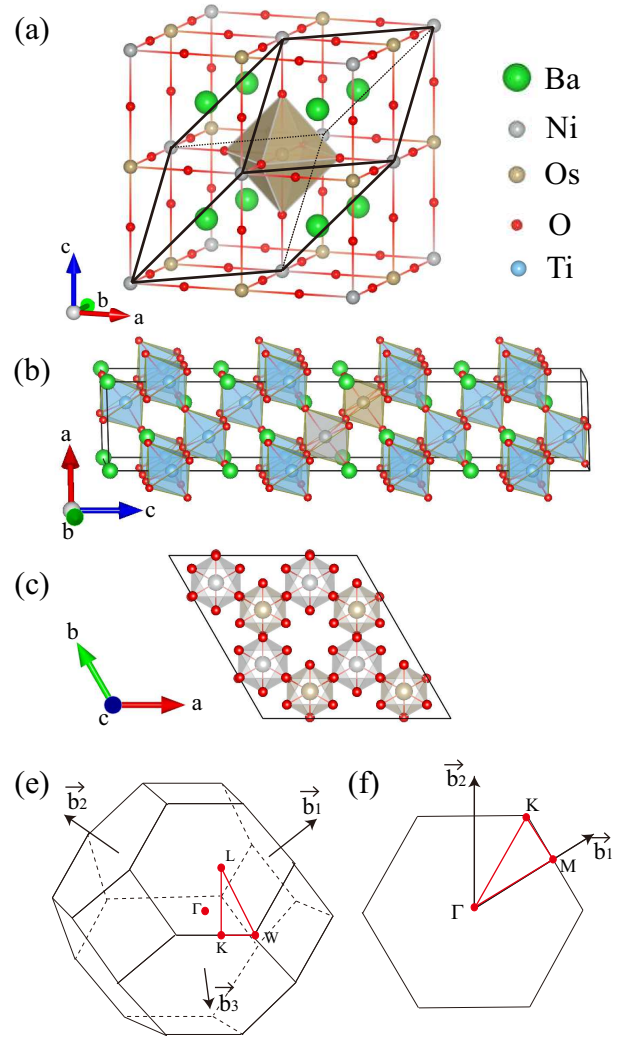}
\caption{(a) Cubic crystal cell of bulk Ba$_2$NiOsO$_6$. Black
lines indicate the fcc primitive unit cell. (b) Side view and (c)
top view of the crystal structure of the (111)
(Ba$_2$NiOsO$_6$)$_{1}$/(Ba$_2$TiO$_3$)$_{10}$ superlattice. In
(c), two different colors denote the X atoms on the two different
planes in the (111) Ba$_2$NiOsO$_6$ monolayer forming a buckled
honeycomb lattice. (e) The Brillouin Zone of bulk Ba$_2$NiOsO$_6$
and (f) the BZ of its (111)
(Ba$_2$NiOsO$_6$)$_{1}$/(Ba$_2$TiO$_3$)$_{10}$ superlattice, with
the basis vectors $\vec{b_{1}}$, $\vec{b_{2}}$ and $\vec{b_{3}}$
of the reciprocal lattices as well as some high-symmetry points
labeled.}
\end{figure}

\begin{figure}
\includegraphics[width=6cm]{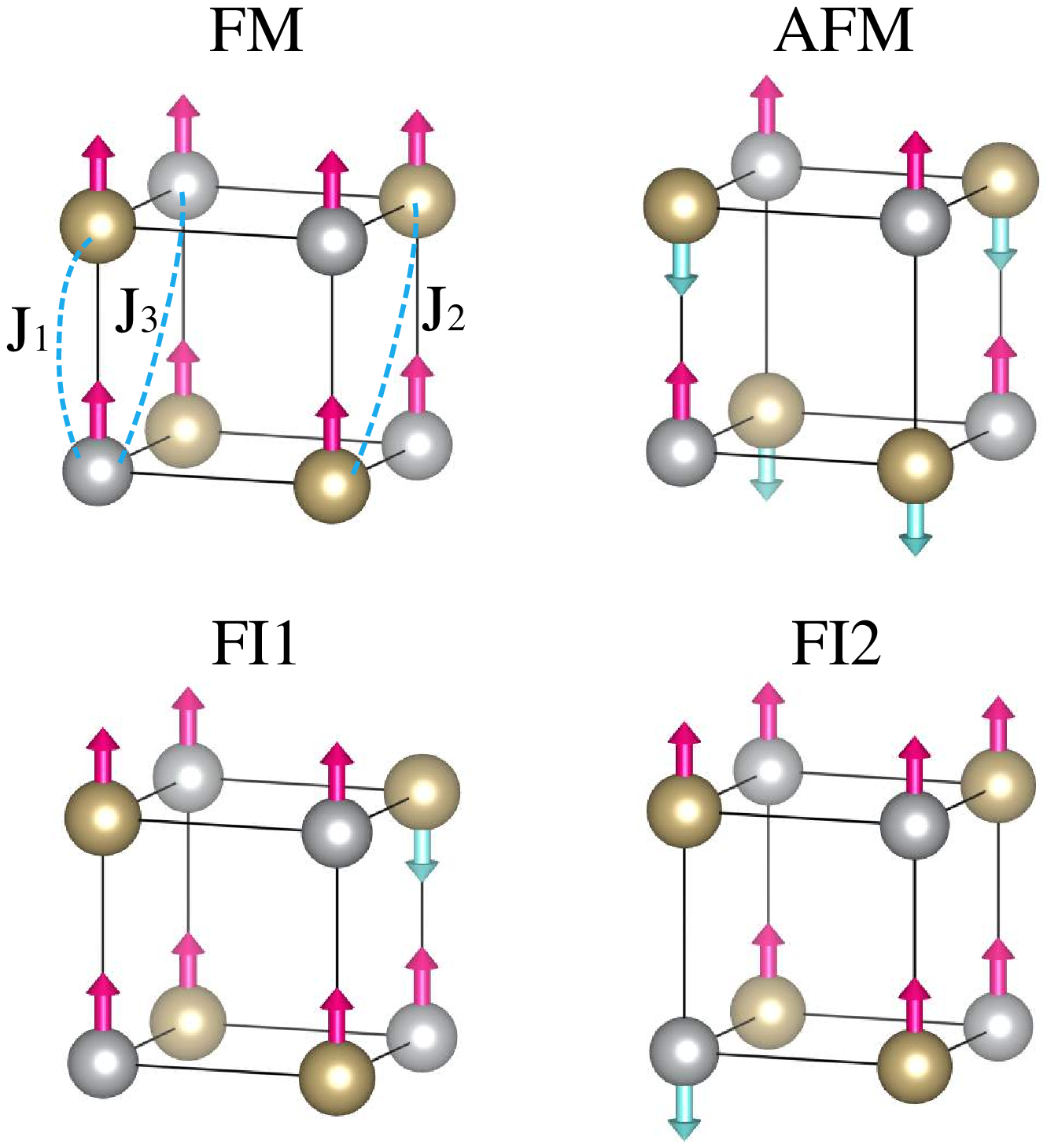}
\caption{(a) Ferromagnetic (FM), (b)
antiferromagnetic (AFM), (c) type I ferrimagnetic (FI1)
and (d) type II ferrimagnetic (FI2) configurations  in bulk Ba$_2$NiOsO$_6$.}
\end{figure}

\begin{figure}
\includegraphics[width=6cm]{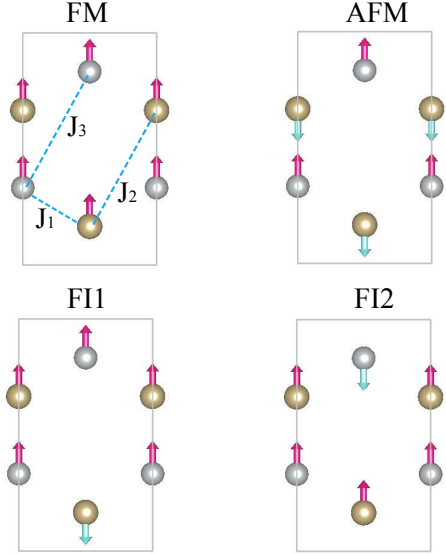}
\caption{(a) Ferromagnetic (FM), (b) antiferromagnetic (AFM), (c)
type I ferrimagnetic (FI1) and (d) type II ferrimagnetic (FI2)
configurations in (111) monolayer Ba$_{2}$NiOsO$_{6}$.}
\end{figure}

For a ferromagnetic solid with at least threefold rotational
symmetry (i.e., tetragonal, trigonal, hexagonal and cubic) and the
magnetization along the rotational $z_{-}$axis, the optical
conductivity tensor can be written as\cite{Wooten72}
\begin{equation}
\bm{\sigma}= \begin{pmatrix} \sigma_{xx}  & \sigma_{xy} & 0 \\
-\sigma_{xy} & \sigma_{yy} & 0 \\
0 & 0 & \sigma_{zz}
\end{pmatrix}.
\end{equation}
Within linear response Kubo formalism~\cite{Wang74},
the absorptive parts of the conductivity tensor elements due to interband transitions are given by
\begin{equation}
\sigma_{1aa} (\omega) = \frac{\alpha}{\omega}
\sum_{i,j}\int_{BZ}\frac{d{\bf k}}{(2\pi)^3}|p_{ij}^{a}|^{2}
\delta(\epsilon_{{\bf k}j}-\epsilon_{{\bf k}i}-\hbar\omega),
\end{equation}
\begin{equation}
\sigma_{2xy} (\omega) = \frac{\alpha}{\omega}
\sum_{i,j}\int_{BZ}\frac{d{\bf k}}{(2\pi)^3}\text{Im}[p_{ij}^{x}p_{ji}^{y}]
\delta(\epsilon_{{\bf k}j}-\epsilon_{{\bf k}i}-\hbar\omega),
\end{equation}
where summations $i$ and $j$ are over the valence and conduction
bands, respectively. $\alpha=\frac{{\pi}{e^{2}}}{{\hbar}{m^{2}}}$
is a material specific constant, $\hbar$$\omega$ is the photon
energy, and $\epsilon_{{\bf k}i}$ is the $i$th band energy at
${\bf k}$ point. Dipole matrix elements $p_{ij}^{a} =
\langle\textbf{k}\emph{j}|\hat{p}_{a}|\textbf{k}i\rangle$ where
$\hat{p}_a$ denotes Cartesian component $a$ of the dipole
operator, are obtained from the band structures within the PAW
formalism\cite{Adolph01}, as implemented in the VASP package. The
integration over the BZ is carried out by using the linear
tetrahedron method (see Ref. [\onlinecite{Temmerman89}] and
references therein). The dispersive parts of the conductivity
tensor elements can be obtained from the corresponding absorptive
parts by use of the Kramer-Kronig transformation~\cite{Bennett65},
\begin{equation}
\sigma{_{2aa}}(\omega) = -
\frac{2\omega}{\pi}P\int_{0}^{\infty}\frac{\sigma_{1aa}(\omega')}{\omega'^{2}-\omega^{2}}d\omega',
\end{equation}
\begin{equation}
\sigma{_{1xy}}(\omega) =
\frac{2}{\pi}P\int_{0}^{\infty}\frac{\omega'\sigma_{2xy}(\omega')}{\omega'^{2}-\omega^{2}}d\omega',
\end{equation}
where $P$ denotes the principal integral.

In the polar geometry, the complex Kerr angle can then be
calculated from the optical conductivity tensor
via~\cite{Guo94,Guo95},
\begin{equation}
\theta_{K} + i\varepsilon_{K} =
\frac{-{\sigma}_{xy}}{{\sigma}_{xx}\sqrt{1+i(4{\pi}/\omega){\sigma}_{xx}}},
\end{equation}
which $\theta$$_{K}$ is the Kerr rotation angle and
$\varepsilon$$_{K}$ the Kerr ellipticity. For a magnetic thin
film, the complex Faraday rotation angle can be written
as~\cite{ravindran1999},
\begin{equation}
\theta _{F}+i\epsilon _{F}=\frac{\omega d}{2c}(n_{+}-n_{-}),
\end{equation}
where $n_+$ and $n_-$ represent the refractive indices for left- and right-handed polarized lights, respectively,
and are related to the corresponding dielectric function (or optical conductivity via expressions
$n_{\pm }^{2}=\varepsilon_{\pm}=1+{\frac{4\pi i}{\omega}}\sigma _{\pm}= 1+{\frac{4\pi i}{\omega}}(\sigma _{xx}\pm i \sigma _{xy})$.
For many magnetic materials, the $\sigma_{xx}$ is generally much larger than the corresponding $\sigma_{xy}$.
Therefore,
$n_{\pm }=[1+{\frac{4\pi i}{\omega}}(\sigma _{xx}\pm i \sigma _{xy})]^{1/2}$
$\approx [1+{\frac{4\pi i}{\omega}}\sigma _{xx}]^{1/2} \mp {\frac{2\pi}{\omega}}(\sigma _{xy}/\sqrt{1+\frac{4\pi i}{\omega}\sigma _{xx}})$.
Consequently,
\begin{equation}
\theta _{F}+i\epsilon _{F}\approx -\frac{2\pi d}{c}\frac{\sigma _{xy}}{\sqrt{1+\frac{4\pi i}{\omega}\sigma _{xx}}}.
\end{equation}

The anomalous Hall conductivity (AHC) is calculated based on the
Berry-phase formalism~\cite{XiaoD10}. Within this Berry-phase
formalism, the AHC ($\sigma_{ij}^{A} = J^c_i/E_j$) is given as a
BZ integration of the Berry curvature for all the occupied
(valence) bands,
\begin{eqnarray}
\sigma_{xy}^{A} = -\frac{e^2}{\hbar}\sum_{n \in VB} \int_{BZ}\frac{d{\bf k}}{(2\pi)^3}\Omega_{xy}^n({\bf k}),\nonumber \\
\Omega_{xy}^n({\bf k}) = -\sum_{n'\neq n}
\frac{2{\rm Im}[p_{ij}^{x}p_{ji}^{y}]}
 {(\epsilon_{{\bf k}n}-\epsilon_{{\bf k}n'})^2},
\end{eqnarray}
where ${\Omega_{ij}^n({\bf k})}$ is the Berry curvature for the
$n$th band at ${\bf k}$. $J^c_i$ is the $i$ component of the
charge current density ${\bf J}^c$ and $E_j$ is the $j$-component
of the electric field ${\bf E}$. Note that the AHC is nothing but
$\sigma$$_{1xy}(\omega)$ in the dc limit, i.e., $\sigma_{xy}^{A} =
\sigma$$_{1xy}(\omega = 0)$. From the Kramers-Kronig relations, we
can obtain a sum rule for $\sigma$$_{1xy}(\omega = 0)$,
\begin{equation}
\sigma{_{1xy}}(\omega = 0) =
\frac{2}{\pi}P\int_{0}^{\infty}\frac{\sigma_{2xy}(\omega')}{\omega'}d\omega'.
\end{equation}
Putting Eq.(3) into this sum rule would result in Eq. (9). Since a
large number of $k$ points are needed to get accurate AHCs, we use
the efficient Wannier interpolation method~\cite{WangX06, LopezMG}
based on maximally localized Wannier functions
(MLWFs)~\cite{MarzariN}. Since the energy bands around the Fermi
level are dominated by Os $t_{2g}$ orbitals, 6 MLWFs per unit cell
of Os $t_{2g}$ orbitals are constructed by fitting to the
GGA+U+SOC band structure in the energy window from -0.69 eV to
2.51 eV for the bulk, and from -0.54 eV to 1.66 eV for the
monolayer. The band structure obtained by the Wannier
interpolation agrees well with that from the GGA+U+SOC
calculation. The AHC ($\sigma_{xy}^A$) was then evaluated by
taking a very dense k-point mesh of  200 $\times 200$ $\times$ 200
and 200 $\times 200$ $\times$ 1 in the BZ for bulk Ba$_2$NiOsO$_6$
and its (111) monolayer, respectively.

\section{Results and discussion}
\subsection{Magnetic properties}

We study four magnetic configurations in both bulk Ba$_2$NiOsO$_6$
and its (111) monolayer, as illustrated in Figs. 2 and 3,
respectively. The calculated total energies of these magnetic
configurations are listed in Table I. It is clear that in both
structures the FM configuration is the ground state. Therefore, we
list in Table II the calculated magnetic moments and band gap of
only the FM state. Table II shows that in bulk Ba$_2$NiOsO$_6$,
both Ni and Os atoms have large spin magnetic moments, being 1.78
$\mu$$_{B}$ and 1.22 $\mu$$_{B}$, respectively. Nevertheless,
because of the hybridization among O $p$, Os $d$, and Ni $d$
orbitals, these spin magnetic moments fall short of 2.0
$\mu$$_{B}$ expected from Ni$^{2+}$ 3$d^8$ ($\emph{t}_{2g}^{6}$
$\emph{e}_{g}^{2}$; $S=1$) and Os$^{6+}$
5$d^{2}$($\emph{t}_{2g}^{2}$ $\emph{e}_{g}^{0}$; $S=1$) ions.
Interestingly, both Ni and especially Os atoms have significant
orbital magnetic moments, being  0.21 $\mu$$_{B}$ and -0.55
$\mu$$_{B}$, respectively. Hund's second rule states that the spin
and orbital moments would be antiparallel if the $d$ shell is less
than half-filled, and otherwise they would be parallel. In
consistence with Hund's second rule, the Ni orbital moment is
parallel to the Ni spin moment while the Os orbital moment is
antiparallel to the Os spin moment. Consequently, because of the
large negative Os orbital moment, the total magnetic moment is
3.37 $\mu$$_{B}$/f.u. in bulk Ba$_2$NiOsO$_6$. This theoretical
value agrees rather well with the total magnetic moment of 3.46
$\mu$$_{B}$/f.u. deduced from the magnetic susceptibility
experiment~\cite{Feng16}. The calculated magnetic moments for the
(111) Ba$_2$NiOsO$_6$ monolayer are similar to that of bulk
Ba$_2$NiOsO$_6$ (see Table II).

\begin{table}\footnotesize
\caption{\label{tab:table1} The properties of bulk and (111)
monolayer Ba$_{2}$NiOsO$_{6}$ from the GGA+U calculations.
\emph{E$_{FM}$}, \emph{E$_{AFM}$}, \emph{E$_{FI1}$}, and
\emph{E$_{FI2}$} denote the total energies of the FM, AFM, FI1,
and FI2 configurations, respectively (see Figs. 2 and 3).
\emph{J$_{1}$} (\emph{d$_{Ni-Os}$}), \emph{J$_{2}$}
(\emph{d$_{Os-Os}$}), and \emph{J$_{3}$} (\emph{d$_{Ni-Ni}$})
represent the nearest Ni-Os, Os-Os, and Ni-Ni exchange coupling
parameters (interatomic distances), respectively. \emph{T$_{c}$}
is the magnetic ordering temperature.}
\begin{ruledtabular}
\begin{tabular}{ccc}
                               &  Bulk       & (111) monolayer    \\ \hline
\emph{E$_{FM}$}  (meV/f.u.)    &   0              &   0    \\
\emph{E$_{AFM}$}  (meV/f.u.)   &   77.80          &  35.57 \\
\emph{E$_{FI1}$} (meV/f.u.)    &   61.89          &  6.635     \\
\emph{E$_{FI2}$} (meV/f.u.)    &   38.20          &  17.57     \\
\emph{d$_{Ni-Os}$} (\AA)       &   4.021          &  4.103    \\
\emph{d$_{Os-Os}$} (\AA)       &   5.687          &  5.696   \\
\emph{d$_{Ni-Ni}$} (\AA)       &   5.687          &  5.696   \\
\emph{J$_{1}$} (meV)           &   6.48           &  5.928    \\
\emph{J$_{2}$} (meV)           &   2.87           & -2.787  \\
\emph{J$_{3}$} (meV)           &  -0.09           & -0.054   \\
\emph{T$_{c}$}   (K)           &$\sim$150 ($\sim$ 100\footnotemark[1])& $\sim$69\\
\end{tabular}
\end{ruledtabular}
\footnotemark[1]{Experimental value from
Ref.~[\onlinecite{Feng16}].}
\end{table}

As mentioned before, using the calculated total energies for the
four magnetic configurations, we evaluate the exchange coupling
parameters between magnetic atoms. The obtained nearest neighbor
Ni-Os ($J_1$), Os-Os ($J_2$), and Ni-Ni ($J_3$) exchange coupling
parameters together with their distances are listed in Table I.
Interestingly, in both systems the magnetic interaction between B
(Ni) and B$'$ (Os) is ferromagnetic and is rather strong. This FM
coupling  between B and B$'$ cations is very rare in double
perovskite oxides~\cite{Jeng03,Wang09}. This explains why the FM
state is the ground state, quite unlike many other double
perovskite oxides in which the AFM is often
favored~\cite{Jeng03,Wang09}. The second near neighbor Os-Os
exchange coupling ($J_2$) is, however, AFM in the monolayer,
although it is still FM in the bulk (Table I). Furthermore, the
second near neighbor Ni-Ni magnetic coupling ($J_3$) is much
smaller than $J_1$. The smallness of $J_3$ could be attributed to
the much localized Ni 3$d$ orbitals in comparison with that of Os
5$d$ orbitals. Note that we also perform the total energy
calculations for the (111) Ba$_2$NiOsO$_6$ monolayer superlattice
in the zigzag-AFM and stripy-AFM configurations (see Figs. 2(c)
and 2(d) in Ref.~\cite{Hslu2018}) to estimate the next-nearest
neighbor Ni-Os coupling parameter (J$^{NN}_{Ni-Os}$). We obtain a
small J$^{NN}_{Ni-Os}$ value of 0.12 meV. This is much smaller
than the nearest neigbor Ni-Os exchange coupling parameter (see J1
in Table I), and thus the J$^{NN}_{Ni-Os}$ values are not listed
in Table I.

Based on the calculated $J_1$ values, we could estimate magnetic
ordering temperature ($T_c$) within a mean-field approximation
given by $ k_BT_c=\frac {1}{3} zJ_1$ where $z$ is the number of
Ni-Os pairs for either Ni or Os atom. Table I shows that such
estimated $T_c$ of 150 K for the bulk agrees quite well with the
experimental value of 100 K~\cite{Feng16}. The estimated $T_c$ of
69 K for the monolayer is smaller than that of the bulk. This
could be expected because of the number of nearest Ni-Os exchange
couplings decreases from six in the bulk to three in the
monolayer.

\begin{table}%[h]
\caption{Spin ($m_s$) and orbital ($m_o$) magnetic moments as well
as band gap ($E_g$) of ferromagnetic Ba$_2$NiOsO$_6$ and its (111)
monolayer from the GGA+U+SOC calculations. The magnetization is
along the $c$-axis.}
\begin{ruledtabular}
\begin{tabular}{ccc}
         & Bulk  &  (111) monolayer \\ \hline

\emph{m$_{s}^{Os}$} ($\mu_B$/atom) & 1.22 & 1.22 \\
\emph{m$_{o}^{Os}$} ($\mu_B$/atom) & -0.55 & -0.50 \\
\emph{m$_{s}^{Ni}$} ($\mu_B$/atom) & 1.78 & 1.75 \\
\emph{m$_{o}^{Ni}$} ($\mu_B$/atom) & 0.21 & 0.21 \\
\emph{m$_{s}^{O}$} ($\mu_B$/f.u.) & 0.61  & 0.61 \\
\emph{m$_{o}^{O}$} ($\mu_B$/f.u.) & -0.08 &-0.14 \\
\emph{m$_{t}^{Os}$} ($\mu_B$/atom) & 0.67 (0.97\footnotemark[1]) & 0.72 \\
\emph{m$_{t}^{Ni}$} ($\mu_B$/atom) & 1.99 (2.13\footnotemark[1]) & 1.96 \\
\emph{m$_{t}^{tot}$} ($\mu_B$/f.u.) & 3.37 (3.46\footnotemark[1]) & 3.78 \\
$E_g$ (eV) & 0.22 (0.31\footnotemark[1]) & 0.37 \\
\end{tabular}
\end{ruledtabular}
\footnotemark[1]{Experimental values from
Ref.~[\onlinecite{Feng16}].} \label{MI}
\end{table}

\subsection{Electronic structure}

\begin{figure}
\includegraphics[width=8cm]{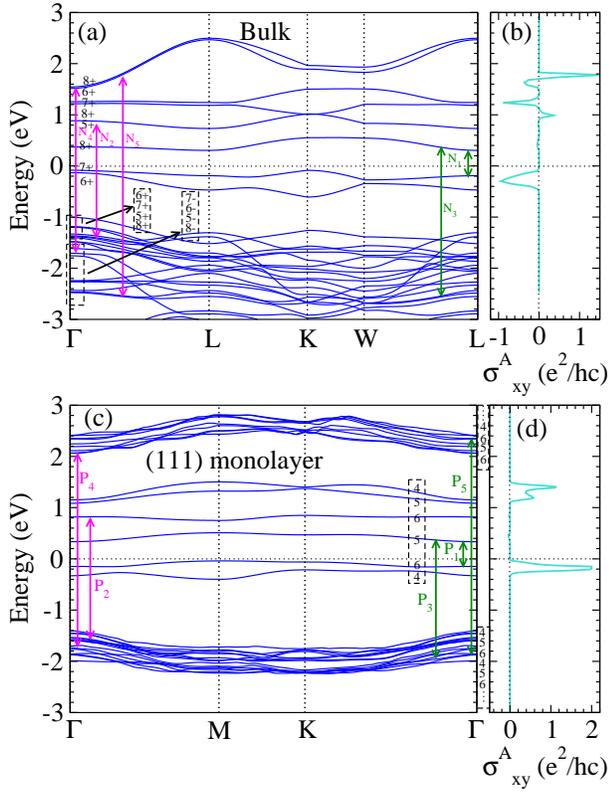}
\caption{(a, c) Relativistic band structure and (b, d) anomalous
Hall conductivity ($\sigma^A_{xy}$) of bulk Ba$_{2}$NiOsO$_{6}$
(upper panels) and its (111) monolayer (lower panels). The FM
magnetization is along the $c$ axis. The zero of energy is placed
at the top of valence bands. In (c), only the Ba2NiOsO6
monolayer-dominated bands are displayed (see the maintext). The
symmetry of band states at the $\Gamma$ point are labeled
according to the irreducible representations listed in Tables IV
and V in Appendix C. The principal inter-band transitions and the
corresponding peaks in the $\sigma_{xy}$ in Figs. 8(c) and 9(c)
are indicated by pink and green arrows.}
\end{figure}

Now let us examine the FM electronic structure of bulk
Ba$_2$NiOsO$_6$ and its (111)
(Ba$_{2}$NiOsO$_{6}$)$_{1}$/(BaTiO$_{3}$)$_{10}$ monolayer
superlattice, which is needed for the following discussion of the
optical conductivity tensor and the magneto-optical effects. The
calculated fully relativistic and scalar-relativistic band
structures are plotted in Figs. 4 and 5, respectively. For clarity
and ease of comparison with bulk, we only show the main
contributions from monolayer Ba$_2$NiOsO$_6$ for (111)
superlattice. Furthermore, the calculated atom- and
orbital-decomposed densities of states (DOSs) for both structures
are displayed in Fig. 6. Figure 4 shows that both structures are a
semiconductor with a small indirect band gap (Table II). Figure 6
indicates that the band gap falls within the spin-up Os 5$d$
$t_{2g}$ dominant bands. In bulk Ba$_2$NiOsO$_6$, the calculated
band gap of 0.22 eV is comparable to the experimental one of
$\sim$ 0.31 eV~\cite{Feng16}. Interestingly, the
scalar-relativistic band structures of bulk Ba$_2$NiOsO$_6$ and
its (111) monolayer are a metal and a semi-metal (Fig. 5),
respectively. When the SOC is included, the $t_{2g}$ (equivalent
to $l=1$) states split into doubly degenerate ($j = 3/2$) occupied
state and nondegenerate ($j = 1/2$) unoccupied state (Fig. 6).
This confirms that the SOC plays an essential role in the
semiconducting gap-opening, and thus Ba$_2$NiOsO$_6$ is known as a
very rare FM Dirac-Mott insulator~\cite{Feng16}.

\begin{figure}
\includegraphics[width=8cm]{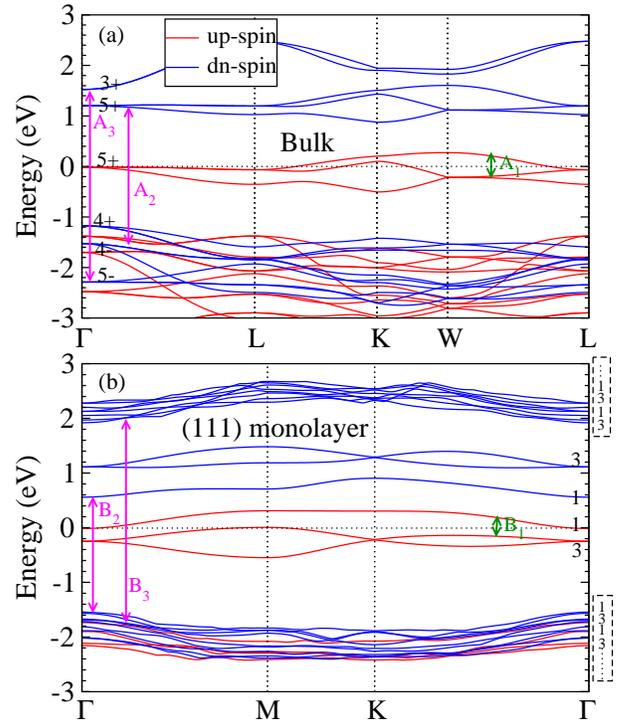}
\caption{Scalar-relativistic band structures of bulk
Ba$_{2}$NiOsO$_{6}$ (a) and its (111) monolayer (b). The zero of
energy is placed at the top of valence bands. In panel (b), only
the Ba$_2$NiOsO$_6$ monolayer dominated bands are displayed (see
the maintext). The symmetry of band states at the $\Gamma$ point
are labeled according to the irreducible representations listed in
Tables IV and V in Appendix C. The principal inter-band
transitions and the corresponding peaks in the $\sigma_{1xx(zz)}$
in Figs. 7(a) and 7(c) are indicated by pink arrows.}
\end{figure}

\begin{figure}
\includegraphics[width=8cm]{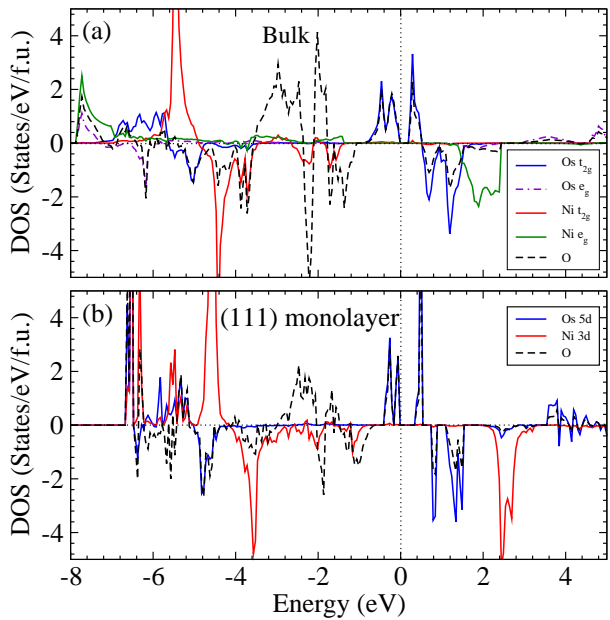}
\caption{Os 5$d$, O, and Ni 3$d$ partial densities of states (DOS)
of bulk Ba$_{2}$NiOsO$_{6}$ (a) and its (111) monolayer (b). The
zero of energy is placed at the top of valence bands.}
\end{figure}

Generally speaking, the DOS of the bulk and the (111) monolayer
are rather similar (see Fig. 6). The energy bands near Fermi level
are predominantly of the Os 5$d$ orbitals. The Os 5$d$ and Ni 3$d$
bands appear in the energy regions of -7.8$\sim$-3.6 eV and
-0.8$\sim$2.4 eV with significant inter-orbital mixing as well as
some small admixture of O 2$p$ states, while the O 2$p$ bands are
mainly located between them. Consequently, because of the strong
Ni 3$d$ - Os 5$d$ hybridization through the O 2$p$ orbital, the Os
5$d$ states are split into bonding and anti-bonding states. The
bonding bands occur in the energy range from -6.8 eV to -4.8 eV,
while the anti-bonding ones are located in the region from -0.8 to
1.6 eV in the vicinity of the Fermi level. However, compared with
the bulk band structure, although the bandwidths of the 6 Os 5$d$
$t_{2g}$-dominated bands near the Fermi level in the monolayer are
only slightly reduced, the bandwidths of the lower valence bands
and upper conduction bands further away from the Fermi level
become noticeably narrowed (see Figs. 4 and 5), due to the reduced
number of neighboring Os atoms in the monolayer. Significantly,
there are bands crossings at K points, forming the so-called Dirac
nodal points in (111) monolayer; In contrast, these band crossings
do not occur at the corresponding W point in bulk Ba$_2$NiOsO$_6$
(Fig. 5). This difference would result in contrasting topological
properties of the two systems, as will be discussed in Sec. III E
below.

In the cubic double perovskite structure, the crystal field at the
transition metal atoms, which sit at the centers of the oxygen
octahedrons and occupy the perovskite B sites alternatively,
should split the $d$ states into two upper energy levels
$e_{g}$($3z^{2}-1$, $x^{2}-y^{2}$) and three bottom energy levels
$t_{2g}$($xy$, $yz$, and $xz$). However, the electronic structure
of Ni 3$d$ in the cubic Ba$_2$NiOsO$_6$ do not follow the theory
of crystal field. In Fig. 5(a), it is clearly that Ni $e_{g}$
states lies lower than Ni $t_{2g}$ states in the up-spin channel.
Additionally, the exchange splitting energy between the up-spin
and down-spin $3d$ $e_{g}$ electrons on the Ni atom is about 9.7
eV, much larger than that of Ni 3$d$ $t_{2g}$ bands of $\sim$ 1.0
eV. This is because that $e_{g}$ orbitals with the wave function
directly pointing to that of O 2$p$ orbitals have much stronger
hybridization with O $2p$ states than that of $t_{2g}$ bands,
which causes the bonding occupied $e_{g}$ states shift to the
lower energy level and the anti-bonding unoccupied $e_{g}$ ones
move toward the higher energy direction. Moreover, it is
interesting to find that the exchange splitting of Os $5d$ (up to
$\sim$ 1.2 eV) states is large, being comparable to the Ni $3d$
band exchange splitting, which is due to the unusual
renormalization of the intra-atomic exchange strength at the Os
sites arising from the Os-Ni interaction, similar to the case of
Sr$_2$FeMoO$_6$~\cite{Sarma000}.

Two different mechanisms of magnetism in double perovskite oxides
have been proposed in the earlier literatures. One is the
hybridization-driven mechanism~\cite{Kanamori001} which leads to a
negative spin polarization at the 4$d$ or 5$d$ site, that is, the
intrinsic spin splitting at the 3$d$ site and an induced spin
splitting at the 4$d$ or 5$d$ site which is oppositely aligned.
However, in Ba$_2$NiOsO$_6$, the Os $5d$ and Ni $3d$ states is
ferromagnetic rather than anti-ferromagnetic coupling. Thus, the
hybridization-driven mechanism is not the origin of the magnetic
coupling between Ni 3$d$ ions and Os 5$d$ ions in Ba$_2$NiOsO$_6$.
The other is the well-known superexchange mechanism based on
Goodenough-Kanamori (G-K) rules~\cite{Goodenough-Kanamori}. In
Ba$_2$NiOsO$_6$, Ni $t_{2g}$ orbitals are completely filled, and
this rules out Os $t_{2g}$- Ni $t_{2g}$ interaction. Although Ni
$e_{g}$ orbitals are half-filled and Os $t_{2g}$ orbitals are
partially filled, they are orthogonal and thus do not contribute
to the magnetic exchange. The remaining superexchange interaction
is between half-filled Ni $e_{g}$ and empty Os $e_{g}$ orbitals,
which should lead to the ferromagnetic coupling. Generally
speaking, being a 5$d$ transition metal, Os has a large
$t_{2g}$-$e_{g}$ crystal-field splitting, thus driving the $e_{g}$
states out of the FM coupling picture in such systems as
Sr$_2$NiOsO$_6$ and Ca$_2$NiOsO$_6$. Nevertheless, Ni $e_{g}$ and
Os $e_{g}$ orbitals could hybridize strongly, as shown clearly in
Fig. 6. It is this interaction between the Ni $e_{g}$ and Os
$e_{g}$ orbitals that leads to the strong ferromagnetic coupling
between the Ni and Os ions in Ba$_2$NiOsO$_6$.

\begin{figure}
\includegraphics[width=8cm]{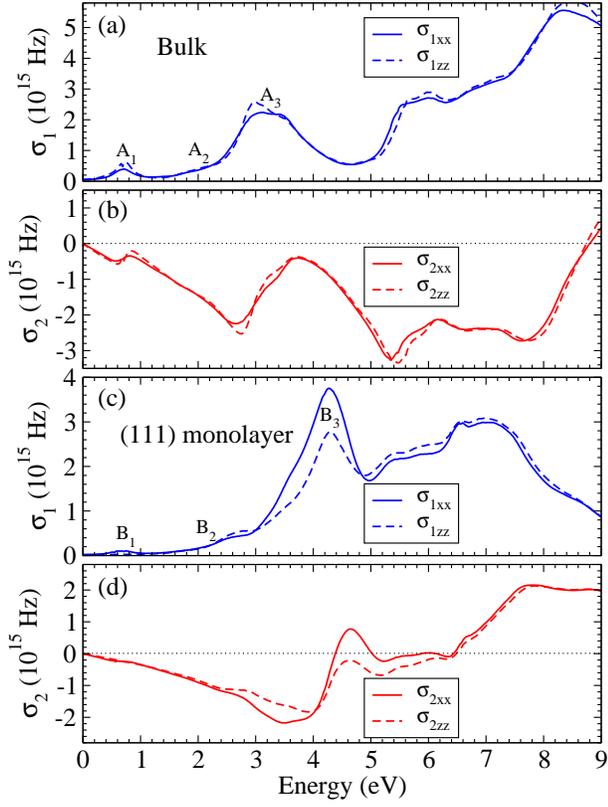}
\caption{(a, c) Real part ($\sigma$$_{1}$) and (b, d) imaginary
part ($\sigma$$_{2}$) of the diagonal elements of the optical
conductivity tensor of bulk Ba$_{2}$NiOsO$_{6}$ (a, b) and its
(111)(Ba$_{2}$NiOsO$_{6}$)$_{1}$/(BaTiO$_{3}$)$_{10}$ monolayer
superlattice (c, d).}
\end{figure}

\begin{figure}
\includegraphics[width=8cm]{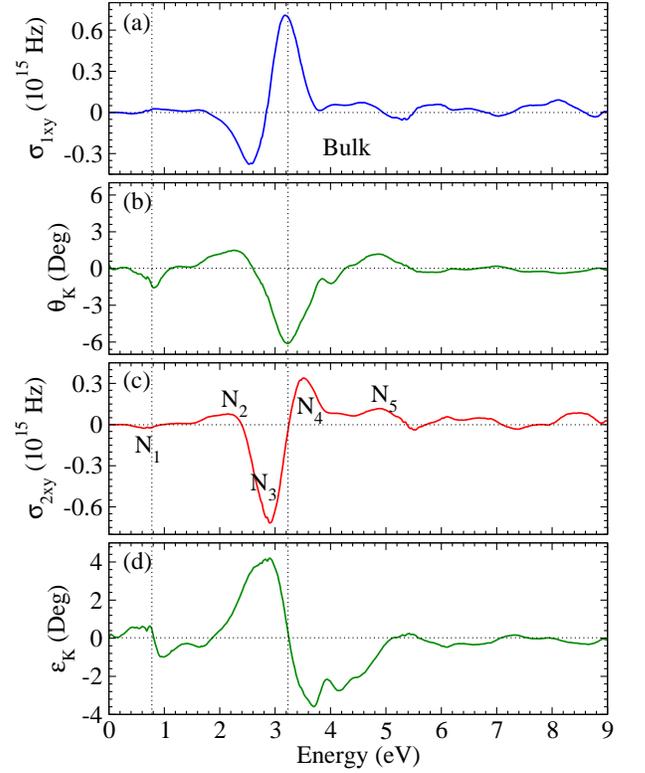}
\caption{(a) Real part ($\sigma$$_{1xy}$) and (c) imaginary part ($\sigma$$_{2xy}$)
of the off-diagonal element of the optical conductivity tensor
as well as (b) Kerr rotation angle ($\theta_K$) and (d) Kerr ellipticity
($\varepsilon_K$) of bulk Ba$_{2}$NiOsO$_{6}$.
}
\end{figure}

\subsection{Optical conductivity}

We calculate the optical conductivity tensors of bulk
Ba$_2$NiOsO$_6$ and its (111) monolayer. The diagonal elements
$\sigma$$_{xx}$ (for in-plane electric field polarization E $\bot$
c) and $\sigma$$_{zz}$ (for out-of-plane electric field
polarization E $\|$ c) of the optical conductivity are displayed
in Fig. 7 for both systems. Overall, the calculated spectra of the
diagonal elements for the different electric field polarizations
in bulk Ba$_2$NiOsO$_6$ are very similar, i.e., this material is
optically isotropic. In particular, they have several identical
peaks. Taking the $\sigma$$_{1xx}$ and $\sigma$$_{1zz}$ spectra as
an example, there are a small peak around 0.7 eV, a prominent twin
peak centered at 3.0 and 3.5 eV, and a broad valley from 5.6 to
8.4 eV. This optical isotropy could be expected from such highly
symmetric crystals as cubic double perovskites. However, for the
(111) superlattic, surprisingly, $\sigma$$_{1xx}$ and
$\sigma$$_{1zz}$ are also similar. The reduced symmetry in the
superlattice causes only small differences. For example, the
prominant B$_3$ peak at $\sim$4.2 eV in the $\sigma$$_{1xx}$
spectrum is only slightly higher than that in the $\sigma$$_{1zz}$
spectrum [see Fig. 7(c)] due to the reduced crystal symmetry.
Nevertheless, compared with bulk case, although the spectral lines
of (111) superlattice are similar at low frequency region, the
peaks in high energy, such as $B_{3}$ peak, are noticeably higher
and narrower. This can be attributed to the noticeably narrowed
bandwidths of the energy bands further away from the Fermi level
[see Fig. 5(b)], as mentioned in the preceding subsection.

The real ($\sigma$$_{1xy}$) and imaginary ($\sigma$$_{2xy}$) parts
of the off-diagonal element of the optical conductivity for bulk
Ba$_2$NiOsO$_6$ are shown in Figs. 8(a) and 8(c), respectively.
These spectra exhibit pronounced oscillatory peaks. Notably, a
large positive peak appears at $\sim$ 3.2 eV in $\sigma$$_{1xy}$,
and that in $\sigma$$_{2xy}$ emerges near 2.0 and 3.5 eV. They
also have a pronounced negative peak at $\sim$ 2.5 eV in
$\sigma$$_{1xy}$ and $\sim$ 2.9 eV in $\sigma$$_{2xy}$. Positive
(negative) $\sigma$$_{2xy}$ suggests that the inter-band
transitions are dominated by the excitations due to the
left-circularly (right-circularly) polarized light. For example,
the negative value in $\sigma$$_{2xy}$ around 2.9 eV suggests that
inter-band transitions induced by right-circularly polarized light
should be stronger. However, the peaks near 2.0 and 3.5 eV
indicate the dominance of inter-band transitions due to
left-circularly polarized light.
%Finally, $\sigma$$_{2xy}$ is
%small or even equal to zero such as 3.2 eV meaning that the
%absorption coefficient for two lights is equal to each other.

Broadly speaking, the $\sigma$$_{1xy}$ and $\sigma$$_{2xy}$ for
the (111) Ba$_2$NiOsO$_6$ monolayer, shown in Figs. 9(a) and 9(c),
respectively, are similar to that of the bulk spectra, and the
positive peak positions such as 3.2 eV in $\sigma$$_{1xy}$, $\sim$
2.0 and 3.5 eV in $\sigma$$_{2xy}$, remain unchanged.
Nevertheless, differences exist for the negative peak positions.
For example, negative peaks appear at $\sim$ 2.2 and $\sim$ 4.1 eV
in $\sigma$$_{1xy}$, and at $\sim$ 2.7 and $\sim$ 4.3 eV in
$\sigma$$_{2xy}$, respectively.

\begin{figure}
\includegraphics[width=8cm]{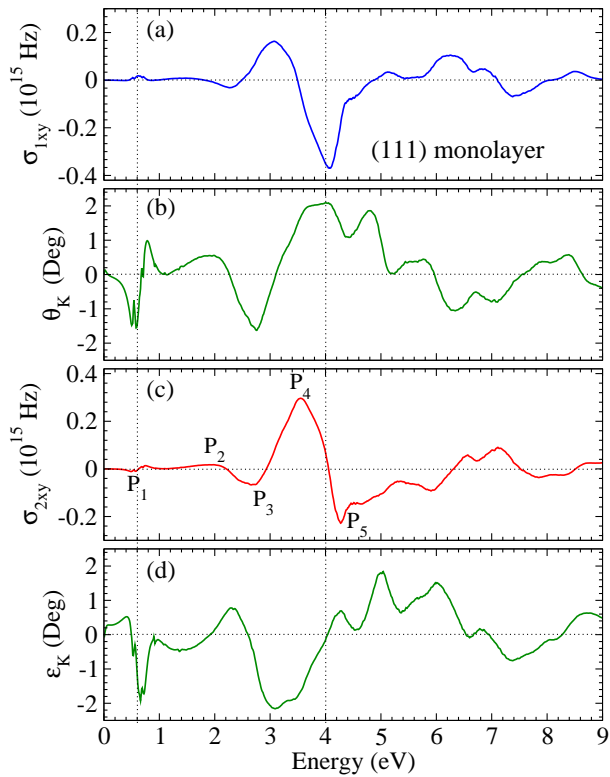}
\caption{(a) Real part ($\sigma$$_{1xy}$) and (c) imaginary part
($\sigma$$_{2xy}$) of the off-diagonal element of the optical
conductivity tensor as well as (b) Kerr rotation angle
($\theta_K$) and (d) Kerr ellipticity ($\varepsilon_K$) of the
(111) (Ba$_{2}$NiOsO$_{6}$)$_{1}$/(BaTiO$_{3}$)$_{10}$ monolayer
superlattice. }
\end{figure}

As Eqs. (2) and (3) suggest, the absorptive parts of the optical
conductivity elements, i.e., $\sigma$$_{1xx}$ and
$\sigma$$_{2xy}$, are directly related to the dipole-allowed
inter-band transitions. This would allow us to understand the
origins of the main peaks in the $\sigma$$_{1xx}$ and
$\sigma$$_{2xy}$ spectra by determining the symmetries of the
calculated band states and the dipole selection rules. The
symmetries of band states at the $\Gamma$-point of the
scalar-relativistic and relativistic band structures of bulk
Ba$_{2}$NiOsO$_{6}$ and its (111) monolayer are displayed in Figs.
4 and 5. Using the dipole selection rules (see Tables VI and VII
in Appendix C), we could assign the main peaks in the
$\sigma$$_{1xx}$ in Fig. 7(a) and 7(c) and the $\sigma$$_{2xy}$ in
Fig. 8(c) and Fig. 9(c) to the inter-band transitions at the
$\Gamma$ point displayed in Figs. 4 and 5 for the two systems.
Taking bulk Ba$_{2}$NiOsO$_{6}$ as an example, we could relate the
A$_{3}$ peak at $\sim$ 3 eV in the $\sigma$$_{1xx}$ [see Fig.
7(a)] to the inter-band transition mainly from the
$\Gamma{_{4}^{-}}$ or $\Gamma{_{5}^{-}}$ state of the down-spin
valence band to the conduction band state $\Gamma{_{5}^{+}}$ or
$\Gamma{_{3}^{+}}$. Of course, in addition to this, there may be
contributions from different inter-band transitions at other $k$
points. Note that without SOC, these band states are doubly
degenerate. When the SOC is included, these band states split [see
Fig. 4(a)], and this results in the magnetic circular dichroism.
Therefore, we could assign the main peaks in the $\sigma$$_{2xy}$
to the principal inter-band transitions at the $\Gamma$-point only
in the relativistic band structure, e.g., displayed in Fig. 4(a).
In particular, we could attribute the pronounced peak N$_{3}$ at
$\sim$ 3.0 eV in the $\sigma$$_{2xy}$ in Fig. 8(c) to the
inter-band transition  from the $\Gamma{_{5}^{-}}$,
$\Gamma{_{6}^{-}}$, $\Gamma{_{7}^{-}}$  or $\Gamma{_{8}^{-}}$
states of valence band to the bottom conduction band state
$\Gamma{_{5}^{+}}$ or $\Gamma{_{8}^{+}}$ shown in Fig. 4(a).

\begin{figure}
\includegraphics[width=8cm]{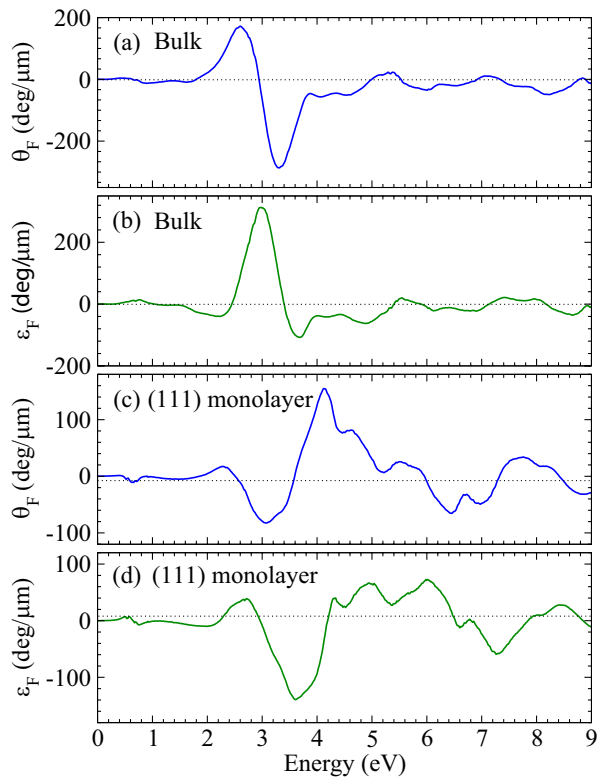}
\caption{(a, c) Faraday rotations ($\theta_F$) and (b, d) Faraday
ellipticities ($\varepsilon_F$) of bulk Ba$_{2}$NiOsO$_{6}$ (upper
two panels) and its (111) monolayer (lower two panels).}
%The calculated real off-diagonal
%$\sigma$$_{1xy}$ (a) and imaginary off-diagonal $\sigma$$_{2xy}$
%(c) components of the optical conductivity tensor, Kerr rotations
%(b) and  Kerr ellipticities (d) for the [001] and [111]
%magnetization directions of FM (111)
%(Ba$_{2}$NiOsO$_{6}$)$_{1}$/(BaTiO$_{3}$)$_{10}$ superlattice.}
\end{figure}

\subsection{Magneto-optical Kerr and Faraday effects}

After examining the electronic, magnetic, and optical properties
of bulk Ba$_2$NiOsO$_6$ and its (111) monolayer, let us now turn
our attention to their magneto-optical Kerr and Faraday effects.
The calculated complex Kerr rotation angles of bulk and (111)
monolayer Ba$_{2}$NiOsO$_{6}$ are displayed in Figs. 8 and 9,
respectively. For bulk Ba$_{2}$NiOsO$_{6}$, the Kerr rotation
angle is remarkably large, reaching up to -1.5{\degree} at $\sim$
0.8 eV, 1.5{\degree} at $\sim$ 2.3 eV and -6{\degree} at $\sim$
3.2 eV. These large values imply that the large MOKE exists in
bulk Ba$_{2}$NiOsO$_{6}$. As discussed already in Sec. I, this
large MOKE stems from the combined effect of the enhanced band
exchange splitting of the Os 5$d$ $t_{2g}$ orbitals caused by the
significant Ni 3$d$ - Os 5$d$ hybridization and the strong SOC of
the Os atoms~\cite{Guo96}. The shape of the Kerr rotation spectrum
for the (111) superlattice, shown in Fig. 9, is similar to that of
bulk Ba$_{2}$NiOsO$_{6}$. The notable difference between the two
systems is the amplitude of the prominent peaks. The Kerr rotation
angles get reduced in the (111) superlatice, mainly because of the
fact that the (111) Ba$_{2}$NiOsO$_{6}$ monolayer has a smaller
density of the magneto-optically active atoms especially Os atoms
than that of the bulk.

Now let us compare the MOKE of the two systems
%bulk and (111) bilayer Ba$_2$NiOsO$_6$
with that of well-known MO materials. The Kerr rotation angles of
most 3$d$ transition metals and their compounds seldom exceed
0.5{\degree} except, e.g., FePt, Co$_2$Pt~\cite{Guo96}, and
PtMnSb~\cite{van Engen83}. Manganese pnictides generally have
excellent MO properties. In particular, MnBi films possess a large
Kerr rotation angle of 2.3{\degree} at 1.84 eV in low
temperatures~\cite{Ravindran999,di1996optical}. The famous MO
material Y$_3$Fe$_5$O$_{12}$ harbors a Kerr rotation of
0.23{\degree} at 2.95 eV. Owing to the strong SOC of 4$d$ and 5$d$
transition metal elements, the large MOKE has also been observed
in half-metallic double perovskites containing 4$d$ and 5$d$
elements. Among these double perovskites, Sr$_2$FeWO$_6$ exhibits
a maximum Kerr rotation of 3.87{\degree}~\cite{vidya2004huge}. On
the whole, the Kerr rotation angles of bulk Ba$_2$NiOsO$_6$ and
its (111) monolayer are at least comparable to these well-known MO
materials.

Figures 8 and 9 show that the Kerr rotation ($\theta_K$) and Kerr
ellipticity ($\varepsilon_K$) spectra in both structures resemble,
respectively, the real part ($\sigma_{1xy}$) and imaginary part
($\sigma_{2xy}$) of the off-diagonal conductivity element except a
reversal of sign. This is not surprising because the Kerr effect
and the off-diagonal conductivity element are connected via Eq.
(6). Indeed, Eq. (6) indicates that the complex Kerr rotation
angle would be linearly related to the $\sigma_{xy}$ if the
longitudinal conductivity ($\sigma_{xx}$) varies smoothly.
%For the few-layer Cr$_2$Ge$_2$Te$_6$, the latter is
%certainly true because here we assume that the substrate is SiO$_2$ with dielectric constant $\varepsilon = 3.9$.
For the photon energy below 1.0 eV, the complex Kerr rotation
angles become unphysically large. This is because the
$\sigma_{xx}$ which is in the denominator of Eq. (6), becomes very
small.
%This explains that in contrast to the few-layer cases, the Kerr rotation of the bulk
%is still visible below 1.2 eV (Fig. 8).

The calculated complex Faraday rotation angles for both bulk and
(111) monolayer Ba$_2$NiOsO$_6$ are displayed in Fig. 10. The
Faraday rotation spectra are rather similar to the corresponding
Kerr rotation spectra as well as the $\sigma_{xy}$ (see Figs. 8
and 9). Figures 7-9 show that the $\sigma_{xx}$ is generally much
larger than the corresponding $\sigma_{xy}$. Therefore, $n_{\pm
}=[1+{\frac{4\pi i}{\omega}}(\sigma _{xx}\pm i \sigma
_{xy})]^{1/2}$ $\approx [1+{\frac{4\pi i}{\omega}}\sigma
_{xx}]^{1/2} \mp {\frac{2\pi}{\omega}}(\sigma
_{xy}/\sqrt{1+\frac{4\pi i}{\omega}\sigma _{xx}})$. Consequently,
$\theta _{F}+i\epsilon _{F}\approx -\frac{2\pi d}{c}(\sigma
_{xy}/\sqrt{1+\frac{4\pi i}{\omega}\sigma _{xx}})$, and this
explains why the complex Faraday rotation more or less follows
$\sigma _{xy}$ (see Figs. 8, 9, and 10).

Remarkably, the maximum Faraday rotation angles are as large as
$\sim$$-250$ deg/$\mu$m at $\sim$ 3.3 eV in bulk Ba$_2$NiOsO$_6$
[see Fig. 10(a)] and $\sim$$160$ deg/$\mu$m at $\sim$ 4.1 eV in
the monolayer [see Fig. 10(c)]. As mentioned above, manganese
pnictides usually have excellent MO properties, and among them
MnBi films possess the largest Faraday rotations of $\sim 80.0$
deg/$\mu$m at 1.77 eV at low
temperatures~\cite{Ravindran999,di1996optical}. However, the
famous MO material Y$_3$Fe$_5$O$_{12}$ possess only a moderate
Faraday rotation of $0.19$ deg/$\mu$m at 2.07 eV. By substituting
Y with Bi, Vertruyen {\it et al.} obtained an enhanced Faraday
rotation of $\sim35.0$ deg/$\mu$m at 2.76 eV in
Bi$_3$Fe$_5$O$_{12}$~\cite{vertruyen08}. Also as mentioned above,
large magneto-optical effects are observed in some half-metallic
double perovskites containing 4$d$ and 5$d$ transition metals. For
example, Sr$_2$FeWO$_6$ possess a large Faraday rotation of $45.0$
deg/$\mu$m~\cite{vidya2004huge}. Clearly, the Faraday rotation
angles for both bulk Ba$_2$NiOsO$_6$ and its (111) monolayer are
larger than these well-known MO materials. Therefore, because of
their excellent MO properties, these Ba$_2$NiOsO$_6$ materials
could find promising applications for, e.g., MO sensors and high
density MO data-storage devices.

\begin{figure}
\includegraphics[width=8cm]{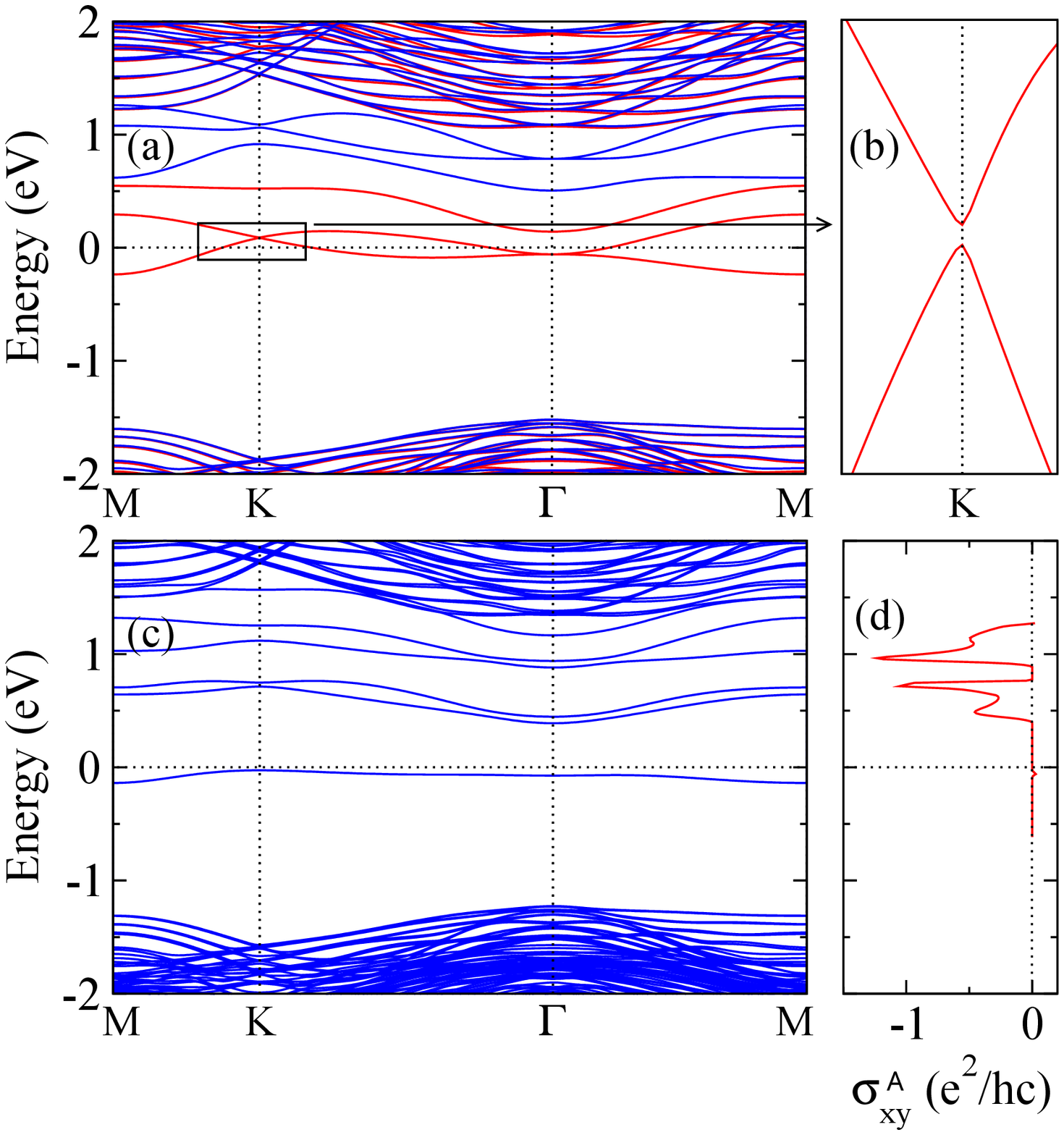}
\caption{(a) Scalar-relativistic and (c) relativistic band
structures as well as (d) anomalous Hall conductivity
($\sigma_{xy}^A$) of the (111) Ba$_{2}$NiReO$_{6}$ monolayer. The part
in the box in (a) is enlarged and displayed in (b). The FM magnetization
is along the $c$-axis and the Fermi level is at 0 eV.}
\end{figure}

\begin{figure}
\includegraphics[width=8cm]{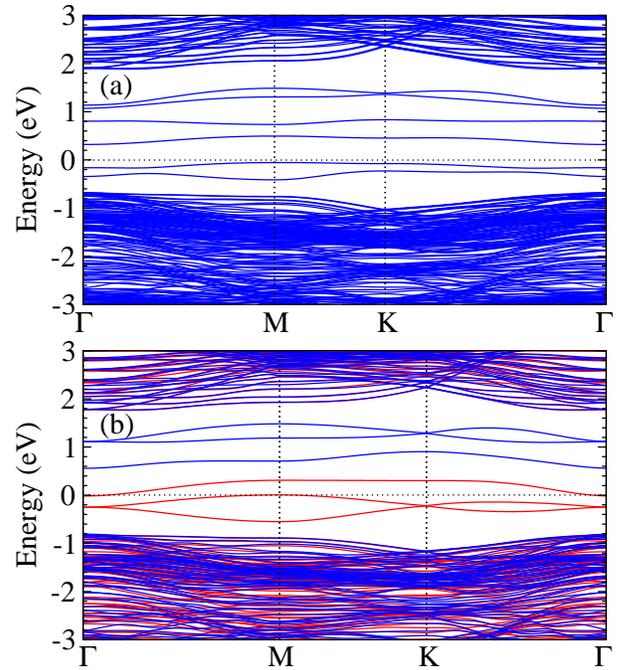}
\caption{(a) Fully-relativistic and (b) scalar-relativistic band
structures of the (111) Ba$_{2}$NiOsO$_{6}$ monolayer. The Fermi
level is at 0 eV. In panel (b), red and blue curves represent
up-spin and down-spin bands, respectively.}
\end{figure}

%\subsection{Topological quantum phases}
\subsection{Anomalous Hall conductivity and topological phases}

As mentioned before, bulk Ba$_2$NiOsO$_6$ and its (111)
superlattice are found to be FM semiconductors when the SOC is
included. We thus could expect that the band gap would be
topologically nontrivial and they could be Chern insulators. To
verify the topological nature of this insulating gap, we calculate
the anomalous Hall conductivity (AHC) ($\sigma_{xy}^A$) for the
two structures. For a three-dimensional (3D) quantum Hall
insulator, $\sigma_{xy}^A$ = $n$ $e^{2}/hc$, where $c$ is the
lattice constant along the $c$ axis normal to the plane of
longitudinal and Hall currents and $n$ is an integer known as the
Chern number ($n_{C}$)~\cite{Halperin87,Zhou2016}. For a normal FM
insulator, however, $\sigma_{xy}^A$ = $0$. The calculated AHC of
bulk Ba$_2$NiOsO$_6$ and its (111) superlattice are displayed in
Figs. 4(b) and 4(d), respectively. Unfortunately, $\sigma_{xy}^A$
is zero within the band gap in both systems, that is, the gaps are
topologically trivial and they are just normal FM insulators.

Here our design principle for engineering topological insulators
is to find a material with its scalar-relativistic band structure
that possesses Dirac points in the BZ, and then examine whether an
energy gap would be opened at these Dirac points when the SOC is
turned-on. For bulk Ba$_2$NiOsO$_6$, an ideal cubic perovskite
structure, the Ni and Os ions sit on a simple cubic lattice with
 the Ni 3$d$ or Os 5$d$ orbitals being split into twofold degenerate $e_{g}$ and threefold degenerate
$t_{2g}$ levels by the octahedral crystal-field. However, such a
lattice geometry usually does not support Dirac points, as one can
see from the calculated scalar-relativistic band structure in Fig.
5(a). This explains that bulk Ba$_2$NiOsO$_6$ remains
topologically trivial when the SOC is switched-on. Recently, Xiao
et al. discovered~\cite{Xiao11} that in a metallic (111) ABO$_{3}$
perovskite bilayer, in which TM B ions form a buckled honeycomb
lattice [see, e.g., Fig. 1(c)], the TM B $e_{g}$ and $t_{2g}$
bands would form Dirac points at the K point in the BZ. As a
result, when the SOC is switched-on, a topologically nontrivial
energy gap would be opened at the Dirac points and hence the
system would be a topological insulator. Indeed, it has been
subsequently demonstrated by many researchers (see, e.g., Refs.
~\cite{Chandra17,Hslu2018,Baidya16} and references therein) that
the topological phase can be achieved in either a (111) simple
perovskite bilayer or a (111) double-perovskite monolayer. In the
(111) Ba$_2$NiOsO$_6$ monolayer, Ni and Os ions form a honeycomb
lattice [see, e.g., Fig. 1(c)], and thus Dirac points appear at
the K point below the Fermi level [Fig. 5(b)]. Therefore, when the
SOC is turned-on, the energy gap opened at the Dirac point is
topologically nontrivial [see Fig. 4(b)]. Nevertheless, we have
also calculated the relativistic band structure for the (001)
Ba$_2$NiOsO$_6$ monolayer. As expected, the (001) monolayer is a
topologically trivial metal.

Interestingly, the calculated $\sigma_{xy}^A$ in (111)
Ba$_2$NiOsO$_6$ superlattice is 2.0 $e^{2}/hc$ within the band gap
opened at these Dirac points with the SOC turned on [Fig. 4(d)],
i.e., the band gap is topologically nontrivial. Therefore, within
the rigid band model, one may speculate that the quantum anomalous
Hall phase would appear in the (111) Ba$_2$NiOsO$_6$ superlattice
when doped with one hole. There are several ways of hole doping
such as chemical substitution~\cite{Richter2017} and electrostatic
gating~\cite{Huang2018,Jiang2018}. Here we explore both the
chemical substitution and electrostatic gating.
%Therefore, based on the crystal structure
%of the Ba$_2$NiOsO$_6$ (111) bilayer with the $d^{8}$-$d^{2}$
%electronic structure,
Specifically, we first consider three kinds of chemical
substitutions, namely, replacing one Ba$^{2+}$ ion with an alkali
metal (X = Li,Na,K) atom as BaXNiOsO$_6$, replacing the Ni$^{2+}$
ion with a transition metal (Y = Sc, Mn, Co, Cu, Ru) atom as
Ba$_2$YOsO$_6$, and also substitution of Re$^{6+}$ for Os$^{6+}$.
Unfortunately, the resultant band structures for the second kind
of substitutions are all metallic. For the first and third kinds
of substitutions, the resultant compounds are semiconductors.
Nonetheless, the semiconducting gaps are all topologically
trivial, i.e., $\sigma_{xy}^A$ is zero within the band gap. As an
example, we display the calculated scalar-relativistic and
relativistic band structures as well as AHC ($\sigma_{xy}^A$) of
the (111) Ba$_2$NiReO$_6$ superlattice in Fig. 11.
%We take the Ba$_{2}$NiReO$_{6}$ (111) bilayer as an
%example, and its scalar-relativistic and relativistic band
%structures as well as anomalous Hall conductivity
%($\sigma_{xy}^A$) are shown in Fig. 11.
Figure 11(d) shows clearly that the $\sigma_{xy}^A$ of the
Ba$_2$NiReO$_6$ superlattice is zero within the semiconducting
gap. We note that the intersection at the K point near the Fermi
level does not exist even without the SOC [Fig. 11 (b)]. This may
explain why the (111) Ba$_2$NiReO$_6$ superlattice is
topologically trivial. We also simulate the hole doping by
electrostatic gating. Here, we perform self-consistent electronic
structure calculations with one less valence electron per f.u.
Unfortunately, the resultant band structure becomes metallic,
indicating that the rigid band model is inapplicable here because
of too strong perturbation due to the hole doping.
%no topological property in these $d^{1}$ systems maybe due to the
%absence of the band crossing at K k-point.

\begin{table*}\footnotesize
\caption{\label{tab:table2} Calculated total and atomic spin (\emph{m$_{s}$}) and orbital (\emph{m$_{o}$}) magnetic moments
%\emph{m$_{s}$} and , orbital magnetic moment \emph{m$_{o}$}, and total
%magnetic moment \emph{m$_{t}$} of Os and Ni atoms
as well as band gap (\emph{E$_{g}$}) of bulk Ba$_2$NiOsO$_6$ as a
function of effective on-site Coulomb repulsions $U_{Os}$ and
$U_{Ni}$ in the GGA + U + SOC method.}
\begin{ruledtabular}
\begin{tabular}{ccccccccccccc}
$U_{Os}$&$U_{Ni}$&\emph{m$_{s}^{Os}$}&\emph{m$_{s}^{Ni}$}&\emph{m$_{s}^{O}$}
&\emph{m$_{o}^{Os}$}&\emph{m$_{o}^{Ni}$}&\emph{m$_{o}^{O}$}&\emph{m$_{s}^{tot}$}&\emph{m$_{o}^{tot}$}&\emph{E$_{g}$}\\
 (eV) & (eV) & ($\mu_B$/atom)&($\mu_B$/atom)&($\mu_B$/f.u.)&($\mu_B$/atom)&($\mu_B$/atom)&($\mu_B$/f.u.)&($\mu_B$/f.u.)&($\mu_B$/f.u.)&(eV)\\
\hline
1 & 3  &1.155 & 1.712  & 0.681 & -0.505 & 0.195 &-0.079 & 3.738 & -0.368 & 0.0 \\
1 & 4  &1.150 & 1.746  & 0.650 & -0.505 & 0.205 &-0.079 & 3.732 & -0.359 & 0.0\\
1 & 5  &1.142 & 1.779  & 0.621 & -0.505 & 0.217 &-0.078 & 3.725 & -0.346 & 0.0\\
2 & 3  &1.232 & 1.710  & 0.672 & -0.547 & 0.196 &-0.085 & 3.771 & -0.413 & 0.2\\
2 & 4  &1.229 & 1.744  & 0.642 & -0.549 & 0.205 &-0.084 & 3.768 & -0.405 & 0.2\\
2 & 5  &1.224 & 1.777  & 0.614 & -0.550 & 0.211 &-0.083 & 3.764 & -0.399 & 0.2\\
3 & 3  &1.314 & 1.708  & 0.655 & -0.570 & 0.197 &-0.086 & 3.793 & -0.435 & 0.4\\
3 & 4  &1.312 & 1.743  & 0.624 & -0.571 & 0.206 &-0.085 & 3.791 & -0.427 & 0.4\\
3 & 5  &1.308 & 1.776  & 0.597 & -0.572 & 0.216 &-0.084 & 3.789 & -0.416 & 0.4\\
\end{tabular}
\end{ruledtabular}
\end{table*}

\section{Conclusions}

In conclusion, by performing systematic first-principles density
functional calculations, we have investigated magnetism,
electronic structure, magneto-optical effects and topological
property of cubic double perovskite Ba$_{2}$NiOsO$_{6}$ and its
(111) (Ba$_{2}$NiOsO$_{6}$)$_{1}$/(BaTiO$_{3}$)$_{10}$ monolayer
superlattice. Interestingly, we find that both structures are rare
FM semiconductors, and the ferromagnetism is driven by strong FM
coupling between neighboring Ni and Os atoms, which in turn arises
from the FM superexchange mechanism due to the abnormally strong
hybridization between half-filled Ni $e_{g}$ and unoccupied Os
$e_{g}$ orbitals. The strong SOC on the Os atom not only opens the
semiconducting gap but also results in a large negative orbital
magnetic moment on the Os atom, thus leading to a total magnetic
moment (3.37 $\mu_B$/f.u.) of less than 4.0
$\mu_B$/f.u.~\cite{Feng16}, expected from the Ni$^{2+}$ 3$d^8$
($\emph{t}_{2g}^{6}$ $\emph{e}_{g}^{2}$; $S=1$) and Os$^{6+}$
5$d^{2}$($\emph{t}_{2g}^{2}$ $\emph{e}_{g}^{0}$; $S=1$) ions. We
also find that because of the enhanced effective intra-atomic
exchange splitting of the Os atoms caused by the Ni 3$d$ - Os 5$d$
hybridization and the strong SOC on the Os sites,
Ba$_{2}$NiOsO$_{6}$ exhibits large MO effects. In particular, the
Kerr and Faraday rotations can be as large as 6$^{\circ}$ and
$250$ deg/$\mu$m, respectively, which are much larger than that of
best-known MO materials. For the (111)
(Ba$_{2}$NiOsO$_{6}$)$_{1}$/(BaTiO$_{3}$)$_{10}$ superlattice, a
large Kerr rotation of $\sim2^{\circ}$ and a large Faraday
rotation of about $160$ deg/$\mu$m is also predicted, although
they are smaller than that bulk Ba$_{2}$NiOsO$_{6}$, mainly due to
the reduced density of magneto-optically active atoms especially
Os atoms in the superlattice. These theoretical findings therefore
suggest that cubic double perovskite Ba$_{2}$NiOsO$_{6}$ and its
(111) superlattice are excellent materials for not only
semiconductor-based spintronics but also magneto-optical devices.
Finally, the calculated AHC reveals that the band gap just below
the Fermi level in the monolayer superlattice is topologically
nontrivial with the gap Chern number of 2 although both structures
are ordinary FM semiconductors. This indicates that the (111)
Ba$_{2}$NiOsO$_{6}$ and related 5$d$ double-perovskite monolayers
may provide an interesting material platform for exploring
magnetic topological phases and phase transitions. This work is
thus expected to stimulate further experimental and theoretical
investigations on these interesting materials.

\begin{acknowledgments}
The authors acknowledge support from the Ministry of Science and
Technology, National Center for Theoretical Sciences, and Academia
Sinica of the Republic of China. H.-S.L. is also supported by the
National Natural Science Foundation of China under Grant No.
11704046.
\end{acknowledgments}

%\clearpage
\section*{Appendix A: GGA+U+SOC calculations}
The calculated spin, orbital, and total magnetic moments for Ni
and Os atoms as well as band gap for bulk Ba$_{2}$NiOsO$_{6}$ from
GGA + U + SOC are listed in Table III. Clearly, the Coulomb
repulsion U on both atoms has little effect on the calculated
magnetic moments. However, the Coulomb repulsion U at Os site
(U$_{Os}$) plays an essential role in opening the band gap, and
the band gap increases with U$_{Os}$. The obtained band gap agrees
well with published experimental value of 0.31 eV when U$_{Os}$
ranging from 2.0 to 3.0 eV. Therefore, the effective Coulomb
repulsions U$_{Os}$ = 2.0 eV and U$_{Ni}$ = 5.0 eV are adopted in
this paper.

\section*{Appendix B: Band structure of (111) Ba$_{2}$NiOsO$_{6}$ monolayer superlattice}
The full band structure of the (111) Ba$_{2}$NiOsO$_{6}$ monolayer
superlattice is displayed in Fig. 12(a) (with SOC) and Fig. 12(b)
(no SOC). The corresponding band structure with only the
monolayer-dominated bands being displayed, is given in Figs. 4(b)
and 5(b), respectively.

\section*{Appendix C: Symmetry analysis}

\begin{table}%\footnotesize
\caption{\label{tab:table3} Symmetry adapted Ni and Os basis functions of the $O_{h}$ point group at
$\Gamma$ for bulk Ba$_2$NiOsO$_6$. $E$ denotes the degeneracy of the band states.}
\begin{ruledtabular}
\begin{tabular}{ccc}
  Symmetry   & $E$ & Basis functions \\
\hline
$\Gamma_{1+}$ ($\Gamma_{1}$, A$_{1g}$) & 1 &$s$  \\
%$\Gamma_{2+}$ ($\Gamma_{2}$, A$_{2g}$) & 1 &$x^{4}(y^{2}-z^{2})+ y^{4}(z^{2}-x^{2})+ z^{4}(x^{2}-y^{2})$  \\
$\Gamma_{3+}$ ($\Gamma_{12}$, E$_{g}$) & 2 &$x^{2}-y^{2}, 2z^{2}-x^{2}-y^{2}$   \\
$\Gamma_{4+}$ ($\Gamma_{15}^{'}$, T$_{1g}$) & 3 &$xy(x^{2}-y^{2}), yz(y^{2}-z^{2}), zx(z^{2}-x^{2})$\\
$\Gamma_{5+}$ ($\Gamma_{25}^{'}$, T$_{2g}$) & 3 &$xy, yz, zx$\\
%$\Gamma_{1-}$ ($\Gamma_{1}^{'}$, A$_{1u}$) & 1 &$xyz[x^{4}(y^{2}-z^{2})+ y^{4}(z^{2}-x^{2})+ z^{4}(x^{2}-y^{2})]$\\
$\Gamma_{2-}$ ($\Gamma_{2}^{'}$, A$_{2u}$) & 1 &$xyz$ \\
%$\Gamma_{3-}$ ($\Gamma_{12}^{'}$, E$_{u}$) & 2 &$xyz(x^{2}-y^{2}), xyz(2z^{2}-x^{2}-y^{2})$ \\
$\Gamma_{4-}$ ($\Gamma_{15}$, T$_{1u}$) & 3 &$x, y, z$ \\
$\Gamma_{5-}$ ($\Gamma_{25}$, T$_{2u}$) & 3 &$z(x^{2}-y^{2}), x(y^{2}-z^{2}), y(z^{2}-x^{2})$
\end{tabular}
\end{ruledtabular}
\end{table}

\begin{table}%\footnotesize
\caption{\label{tab:table4} Symmetry adapted Ni and Os basis functions of the $C_{3v}$ point group at
$\Gamma$ for the (Ba$_2$NiOsO$_6$)$_1$/(BaTiO$_3$)$_{10}$ superlattice. $E$ denotes the degeneracy of the band states.}
\begin{ruledtabular}
\begin{tabular}{ccccccccccccc}
 Symmetry & $E$ & Basis functions \\
\hline
$\Gamma_{{1}}$ (A$_{1}$) & 1 &$z$; $x^{2}+y^{2}$; $z^{2}$\\
$\Gamma_{{2}}$ (A$_{2}$) & 1 & $R_{z}$  \\
$\Gamma_{{3}}$ (E) & 2 &$x, y$; $x^{2}-y^{2}, xy$; $xz, yz$; $R_{x}, R_{y}$
\end{tabular}
\end{ruledtabular}
\end{table}

\begin{table}%\footnotesize
\caption{\label{tab:table3} Dipole selection rules between the band states of point group $O_{h}$ at $\Gamma$.
Relationships between the single group and double group representations are:
$\Gamma_1^+\rightarrow\Gamma_6^+$; $\Gamma_3^+\rightarrow \Gamma_5^+ , \Gamma_8^+$;
$\Gamma_4^+\rightarrow\Gamma_5^+ , \Gamma_6^+ , \Gamma_7^+ , \Gamma_8^+$;
$\Gamma_5^+\rightarrow\Gamma_6^+ , \Gamma_7^+ , \Gamma_8^+$;
$\Gamma_2^-\rightarrow\Gamma_7^-$;
$\Gamma_4^-\rightarrow\Gamma_5^- , \Gamma_6^- , \Gamma_7^- , \Gamma_8^+$;
$\Gamma_4^-\rightarrow\Gamma_5^- , \Gamma_6^- , \Gamma_7^- , \Gamma_8^+$.
\cite{Eberhardt}}
\begin{ruledtabular}
\begin{tabular}{cc}
       & $E\perp c$ $\&$ $E\parallel c$ \\ \hline
Single group& $\Gamma_{1+} \longleftrightarrow  \Gamma_{4-} $ \\
%           & $\Gamma_{2+} \longleftrightarrow  \Gamma_{5-} $ \\
            & $\Gamma_{3+} \longleftrightarrow  \Gamma_{4-}, \Gamma_{5-}$ \\
            & $\Gamma_{4+} \longleftrightarrow  \Gamma_{1-}, \Gamma_{3-}, \Gamma_{4-}, \Gamma_{5-}$ \\
            & $\Gamma_{5+} \longleftrightarrow  \Gamma_{2-}, \Gamma_{3-}, \Gamma_{4-}, \Gamma_{5-}$ \\
%           & $\Gamma_{1-} \longleftrightarrow  \Gamma_{4+} $ \\
            & $\Gamma_{2-} \longleftrightarrow  \Gamma_{5+} $ \\
%           & $\Gamma_{3-} \longleftrightarrow  \Gamma_{4+}, \Gamma_{5+} $\\
            & $\Gamma_{4-} \longleftrightarrow  \Gamma_{1+}, \Gamma_{3+},\Gamma_{4+}, \Gamma_{5+}$ \\
            & $\Gamma_{5-} \longleftrightarrow  \Gamma_{2+}, \Gamma_{3+},\Gamma_{4+}, \Gamma_{5+}$ \\
Double group& $\Gamma_{1+} \longleftrightarrow  \Gamma_{4-} $ \\
%           & $\Gamma_{2+} \longleftrightarrow  \Gamma_{5-} $ \\
            & $\Gamma_{3+} \longleftrightarrow  \Gamma_{4-}, \Gamma_{5-}$ \\
            & $\Gamma_{4+} \longleftrightarrow  \Gamma_{1-}, \Gamma_{3-}, \Gamma_{4-}, \Gamma_{5-}$ \\
            & $\Gamma_{5+} \longleftrightarrow  \Gamma_{2-}, \Gamma_{3-}, \Gamma_{4-}, \Gamma_{5-}$ \\
%           & $\Gamma_{1-} \longleftrightarrow  \Gamma_{4+} $ \\
            & $\Gamma_{2-} \longleftrightarrow  \Gamma_{5+} $ \\
%           & $\Gamma_{3-} \longleftrightarrow  \Gamma_{4+}, \Gamma_{5+} $\\
            & $\Gamma_{4-} \longleftrightarrow  \Gamma_{1+}, \Gamma_{3+},\Gamma_{4+}, \Gamma_{5+}$ \\
            & $\Gamma_{5-} \longleftrightarrow  \Gamma_{2+}, \Gamma_{3+},\Gamma_{4+}, \Gamma_{5+}$
\end{tabular}
\end{ruledtabular}
\end{table}

Bulk Ba$_2$NiOsO$_6$ has the space group $Fm\bar{3}m$ with its
point group $O_{h}$ having 48 symmetry operations. The
site-symmetry point group for Ni and Os atoms is also $O_{h}$.
(111)(Ba$_2$NiOsO$_6$)$_1$/(BaTiO$_3$)$_{10}$ monolayer
superlattice, however, has the space group $P3ml$ with its point
group $C_{3v}$ including six symmetry operations. The
site-symmetry point group for Ni and Os atoms is $C_{3v}$. To
determine the symmetry of the band states at the center
($\Gamma$-point) of the Brillouin zone (BZ), the symmetry adapted
basis functions, formed from the atomic orbitals localized at the
Ni and Os sites, are derived using the projection method of group
theory~\cite{Dresselhaus09}, as listed in Tables V and VI for the
$O_{h}$ and $C_{3v}$ point groups, respectively. By comparing the
calculated orbital characters of the band states at the
$\Gamma$-point with the symmetry-adapted basis functions (Tables V
and VI), one can determine the symmetries of the $\Gamma$-point
band states for bulk Ba$_2$NiOsO$_6$ and its
(Ba$_2$NiOsO$_6$)$_1$/(BaTiO$_3$)$_{10}$ superlattice, as shown in
Figs. 4 and 5.

\begin{table}%\footnotesize
\caption{\label{tab:table4} Dipole selection rules between the band states of point group $C_{3v}$ at $\Gamma$.
Relationships between the single group and double group representations are:
$\Gamma_1\rightarrow\Gamma_5, \Gamma_6$; $\Gamma_1\rightarrow\Gamma_5, \Gamma_6$;
$ \Gamma_3\rightarrow \Gamma_4, \Gamma_5, \Gamma_6$.\cite{Eberhardt}}
\begin{ruledtabular}
\begin{tabular}{ccc}
         & $E\perp c$ & $E\parallel c$ \\ \hline
Single group  & $\Gamma_{1} \longleftrightarrow \Gamma_{3}$ & $\Gamma_{1} \longleftrightarrow \Gamma_{1}$ \\
              & $\Gamma_{2} \longleftrightarrow \Gamma_{3}$ & $\Gamma_{2} \longleftrightarrow \Gamma_{2}$ \\
     & $\Gamma_{3} \longleftrightarrow \Gamma_{1}, \Gamma_{2}, \Gamma_{3} $ & $\Gamma_{3} \longleftrightarrow \Gamma_{3}$ \\
double group  & $\Gamma_{1} \longleftrightarrow \Gamma_{3}$ & $\Gamma_{1} \longleftrightarrow \Gamma_{1}$ \\
              & $\Gamma_{2} \longleftrightarrow \Gamma_{3}$ & $\Gamma_{2} \longleftrightarrow \Gamma_{2}$ \\
     & $\Gamma_{3} \longleftrightarrow \Gamma_{1}, \Gamma_{2}, \Gamma_{3} $ & $\Gamma_{3} \longleftrightarrow \Gamma_{3}$
\end{tabular}
\end{ruledtabular}
\end{table}

Given the known symmetries of the band states at a $k$-point in
the BZ, the possible direct inter-band transitions can be worked
out using the dipole selection rules. The dipole selection rules
for the $O_{h}$ and $C_{3v}$ point groups\cite{Eberhardt} are
listed in Tables VI and VII, respectively.
%Since $C_{3i}$ is a subgroup of $D_{3d}$, we deduce the dipole selection rules
%for $C_{3i}$ (Table S3) from that of $D_{3d}$ reported in ~\cite{Murray72}.
Using these selection rules, we assign the prominent peaks in the
optical conductivity (Figs. 7, 8 and 9) of bulk Ba$_2$NiOsO$_6$
and its (Ba$_2$NiOsO$_6$)$_1$/(BaTiO$_3$)$_{10}$ superlattice to
the principal inter-band transitions at the $\Gamma$ point, as
shown in Figs. 4 and 5.

\end{document}